\documentclass[iop,revtex4]{emulateapj}

\slugcomment{Accepted for publication in The Astrophysical Journal, February 12 2014}

\newcommand{\ind}[1]{_{\mathrm{#1}}}

\shorttitle{Red Giants in Eclipsing Binaries}
\shortauthors{Gaulme et al.}

\begin{document}


\title{Surface Activity and Oscillation Amplitudes of Red Giants in Eclipsing Binaries}


\author{P. Gaulme}
\affil{Department of Astronomy, New Mexico State University, P.O. Box 30001, MSC 4500, Las Cruces, NM 88003-8001, USA}
\affil{Apache Point Observatory, 2001 Apache Point Road, P.O. Box 59, Sunspot, NM 88349, USA}
\email{gaulme@nmsu.edu}

\author{J. Jackiewicz}
\affil{Department of Astronomy, New Mexico State University, P.O. Box 30001, MSC 4500, Las Cruces, NM 88003-8001, USA}

\author{T. Appourchaux}
\affil{Institut d'Astrophysique Spatiale,
Universit\'e Paris-Sud 11 and CNRS (UMR 8617), B\^{a}timent 121, F-91405 Orsay cedex, France}

\and
\author{B. Mosser}
\affil{LESIA, CNRS, Universit\'e Pierre et Marie Curie,
Universit\'e Denis Diderot, Observatoire de Paris, 92195 Meudon
cedex, France}




\begin{abstract}

Among 19 red-giant stars belonging to eclipsing binary systems that have been identified in \textit{Kepler} data, 15 display solar-like oscillations. We study whether the absence of mode detection in the remaining 4 is an observational bias or  possibly  evidence of mode damping that originates from tidal interactions. A careful analysis of the corresponding \textit{Kepler} light curves shows that modes with amplitudes that are usually observed in red giants would have been detected if they were present. We observe that mode depletion is strongly associated with short-period systems, in which stellar radii account for 16-24\,\% of the semi-major axis, and where red-giant surface activity is detected.  We suggest that when the rotational and orbital periods synchronize in close binaries, the red-giant component is spun up, so that a dynamo mechanism starts and generates a magnetic field, leading to observable stellar activity. Pressure modes would then be damped as acoustic waves dissipate in these fields.

\end{abstract}


\keywords{stars: oscillations --- binaries: eclipsing --- stars: evolution}

\section{Oscillating Red Giants in Eclipsing Binaries}
Binary systems hosting at least one star with detectable solar-like pulsations are becoming important astrophysical targets of particular interest because they provide a way to calibrate asteroseismology. As illustrated by the CoRoT and \textit{Kepler} missions \citep{Baglin_2009,Borucki_2010}, asteroseismology is an efficient method to measure masses and radii of large numbers of stars, which is of prime importance to test stellar evolution. However, a crucial test of direct comparison of  asteroseismic masses and radii of a large sample of stars with independent measurements of those quantites \citep[e.g.][and refs.~therein]{Gaulme_2013b} has not yet been carried out.  Eclipsing binary systems (hereafter EBs) potentially permit such an exercise by allowing for accurate determination of masses and radii of both stars by combining photometric and radial-velocity measurements, provided that spectral lines are detectable for both components. 

So far, all the stars known to both display acoustic modes and belong to EBs are red-giants (hereafter RGs), and all have been detected by the \textit{Kepler} mission. The first detection was the 408-day period system KIC 8410637, whose mass was recently determined with radial-velocity measurements and agrees well with the asteroseismic result \citep{Hekker_2010,Frandsen_2013}.
Since then, \citet{Gaulme_2013} reported a list of 17 \textit{bona fide} new RG eclipsing-binary candidates (hereafter RGEBs), of which 12 displayed oscillations. Their orbital periods range from 15 to 987 days, their eccentricities from 0  to 0.7, and their companions are main-sequence stars (from spectral type L/M to F), except one that is  a second RG.  To be used as asteroseismic calibrators, the orbital periods of RGEBs must be measured with high precision, which implies that orbits be shorter than observation runs, i.e. $\approx1500$~days for \textit{Kepler}. More recently, \citet{Beck_2013} reported the discovery of 18 new ``heartbeat'' stars, which are eccentric binaries undergoing dynamic tidal distortions \citep{Welsh_2011}. Each of the 18 systems has a RG component displaying solar-like oscillations. The physical parameters of these systems were derived from modeling \textit{Kepler}'s light curves and radial-velocity measurements.

Eclipsing binaries are also interesting for the physical processes resulting from tidal interactions, which may influence their evolution when the stars are close enough.  In this paper, we find evidence of a puzzling influence that binarity has on mode amplitudes and the surface activity of RGs. We consider all the known RGEBs, i.e. KIC 8410637, those from \citet{Gaulme_2013}, and two new candidates. 
We aim to explore why, among the 19 systems, only 15 display solar-like oscillations. The peculiarity of the four cases with no RG pulsations is their rather short orbital periods ($\leq 41$ days) and high quasi-periodic photometric variability (17 to 27\,\% in relative flux) indicative of large starspots. Taken together, this suggests that the absence of modes has four possible reasons: the modes are buried in the noise related to photometric variations, the mode frequencies are above the Nyquist frequency,  modes are damped because of activity due to strong binarity effects, or these RGs  are simply not pulsating. Damped modes would be consistent with \citet{Chaplin_2011b}, who show that stellar activity inhibits the amplitude of solar-like oscillations.

We first present an update of the RGEB catalog by making use of about four years of \textit{Kepler} data (quarters 0 to 16) to perform eclipse modeling and to estimate global asteroseismic parameters (Sec.~\ref{sect_update}). Then we address the question of mode amplitudes and show evidence that mode damping must be occurring in conjunction with the presence of contrast features (spots, faculae) on the surface of the RGs that are tidally locked in phase (Sect. \ref{sect_obs} \& \ref{sect_discussion}).

\section{Update of the RGEB list}
\label{sect_update}
\subsection{Methods}

The specific data processing required for this analysis entails modeling eclipse shapes by removing stellar activity, and measuring solar-like oscillations by removing eclipse features. The methods used here are an extension of those developed in \citet{Gaulme_2013} and are detailed further in the  Appendix. We specifically work with the RAW public data that are available on the \textit{Kepler} MAST website.

\begin{figure}
\includegraphics[width=0.45\textwidth]{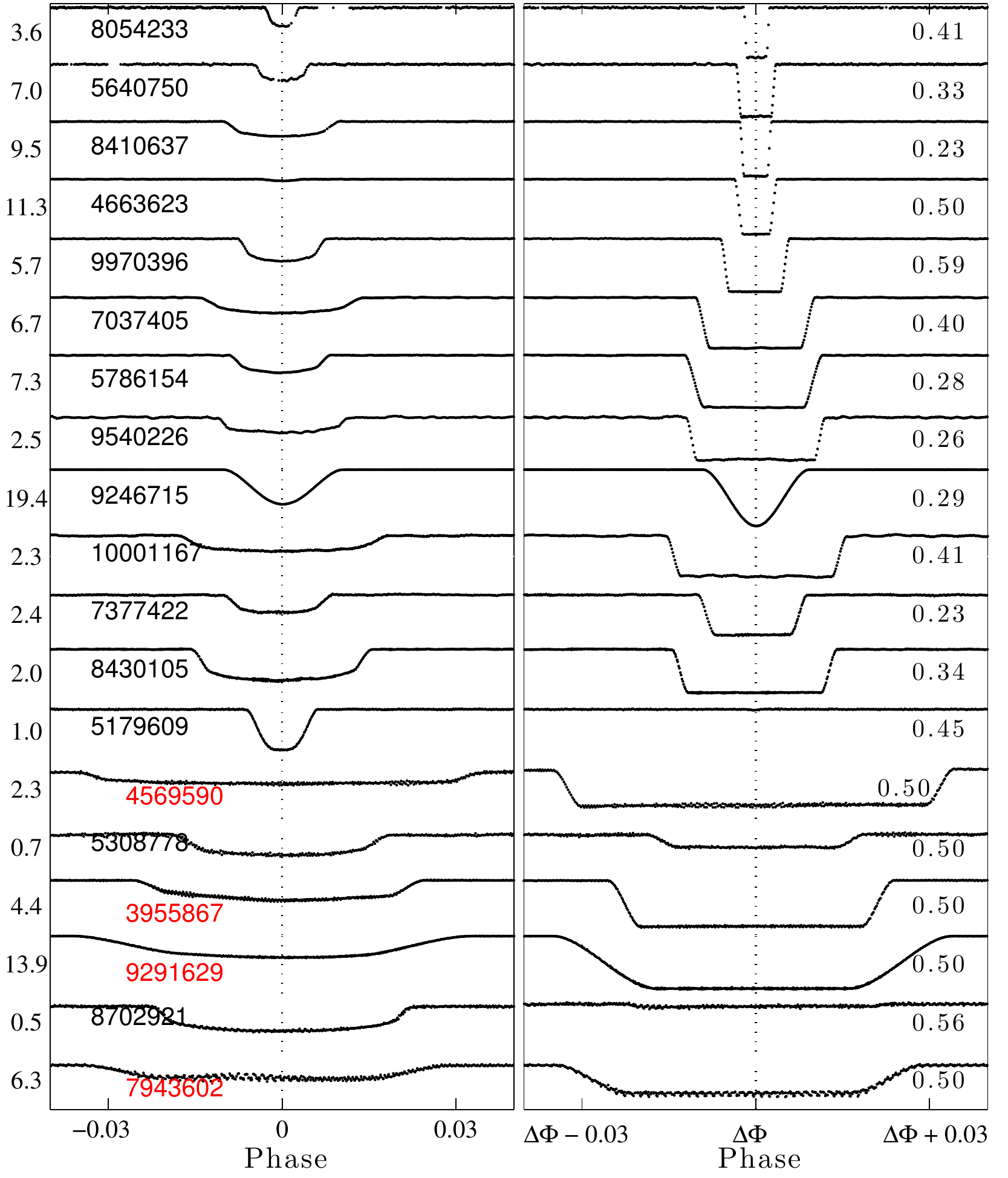}
  \caption{Folded light curves for the 15  eclipsing binaries with a RG displaying solar-like oscillations and for the four longest-period RGEBs with no RG oscillations \citep{Gaulme_2013}.  They are sorted by decreasing orbital period from top to bottom. In each case, the primary eclipse (secondary star eclipsing the RG) has been set to a phase of zero (left panels). The secondary eclipses (red giant eclipsing the secondary star, right panels) have also been aligned, and the given $\Delta\Phi$ value (right panels) indicates the phase of the secondary eclipse with respect to the primary (an eclipsing binary with a circular orbit would have $\Delta\Phi = 0.5$). The $y$-axis label indicates the dimming of the normalized relative flux in percent, corresponding to the deepest eclipse (primary or secondary).  Each system is labeled by its \textit{Kepler} ID number, and labels are shifted to right and printed in red (online version) for those where no modes are detected. \label{fig_LC_ecl}}
\end{figure}

Eclipse modeling consists of retrieving physical parameters of a binary system from the eclipse duration, depth, and shape.  These systems are composed of an RG and a companion  that is usually a main-sequence star. Light curves folded on the orbital period for all systems are shown in Fig.~\ref{fig_LC_ecl}. We define the primary eclipse to be that where the companion star passes in front of the RG, and vice versa for the secondary. For cases composed of a sun-like main-sequence star and a cooler but larger RG, secondary eclipses are deeper than primary ones because of the  temperature ratio. Regarding their shapes, primary eclipses are dominated by the RG limb-darkening function, whereas secondary eclipses are flat except during ingress and egress. Accurate modeling leads to estimations of the ratio of stellar radii to semi-major axis $R_1/a$ and $R_2/a$, the ratio of effective temperatures $T_2/T_1$, the limb-darkening parameters of each star, the orbital period $P$ and eccentricity $e$, the inclination of the orbital plane $i$, and the argument of periastron. Sometimes it is not possible to retrieve all of the parameters, particularly in cases where eclipses are grazing and thus shallow, because the majority of these parameters are degenerate. Radial-velocity measurements help in constraining binary systems by providing independent estimates of $e$ and $a$, as well as masses. Radial-velocity acquisition for the targets considered in this paper is in progress with the 3.5m telescope at the Apache Point Observatory. We carry out eclipse modeling to get orbital periods and insights on the nature of the RG companions, as well as rather rough estimates of their radius, mass, and effective temperature. The modeling is performed with the JKTEBOP code \citep{Southworth_2009}, and main results are given in Table~\ref{tab_1}.

\begin{figure}
\includegraphics[width=0.45\textwidth]{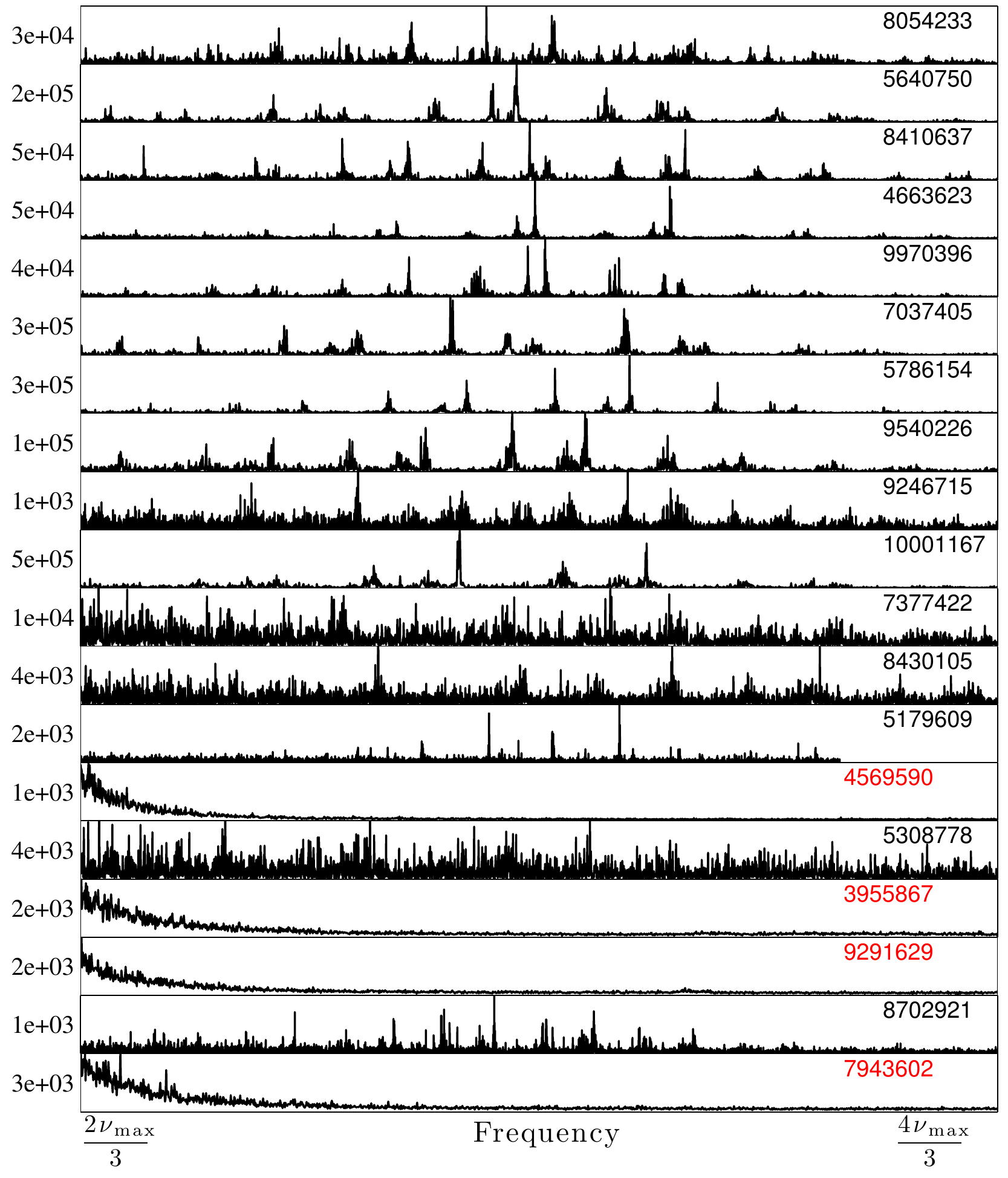}
  \caption{Power spectral density (PSD) of the 15 light curves in which solar-like oscillations are detected and the four where they are not detected, sorted by decreasing orbital period from top to bottom. All PSD are plotted from $2\nu\ind{max}/3$ to $4\nu\ind{max}/3$. In the 4 stars where no modes are detected, the PSD are plotted from 30~$\mu$Hz to the Nyquist frequency (283~$\mu$Hz). Each system is labeled by its \textit{Kepler} ID number. This label is shifted to the left and printed in red (online version) for those where no modes are detected. The $y$-axis indicates the range in power in ppm$^2$\,$\mu$Hz$^{-1}$ for each individual star. \label{fig_PDS}}
\end{figure}

We extract the global asteroseismic parameters from the  power density spectra of the time series, which are shown in Fig.~\ref{fig_PDS}. The frequency at maximum amplitude of solar-like oscillations $\nu\ind{max}$ is measured by fitting the mode envelope with a Gaussian function and the background stellar activity with a sum of two semi-Lorentzians. The large frequency separation $\Delta\nu$ is obtained from the filtered autocorrelation of the time series \citep{Mosser_Appourchaux_2009}. According to the asteroseismic scaling relations for RG stars \citep{Mosser_2013}, we retrieve masses $M_1$ and radii $R_1$  by using effective temperatures from \citet{Huber_2013}. The evolutionary stage of RGs - asymptotic giant branch (AGB), red-giant branch (RGB), main red clump (RC), secondary red clump (RC2) - may be deduced from the study of mixed modes measured in the power spectra. When mixed modes are not detected it is still possible to infer the nature of an RG based on $\Delta\nu$, or on mass and radii criteria \citep{Mosser_2012}. Moreover, as suggested by \citet{Beck_2013}, an RC or AGB identification can be discarded for a star with mass lower than $1.8\ M_\odot$ if the separation between companions is less than 200~$R_\odot$ along its orbit, because it corresponds to the size that a low-mass star reaches at the tip of the RGB. The engulfment of the RG's companion would occur before reaching the RC stage.

Combining outputs from eclipse modeling and asteroseismic measurements leads to the semi-major axis $a$ and the companion star radius $R_2$. Then, from the RG asteroseismic mass, semi-major axis, and Kepler's third law, we get a proxy of the companion's mass, whose precision is at best  20\,\%. In this way, as mentioned in \citet{Gaulme_2013},  masses of companion stars appear systematically overestimated. The derivation of $T_2$ rests entirely on eclipse modeling and the assumption that the \citet{Huber_2013} value is the correct RG temperature. The results of this combined analysis are provided in Table~\ref{tab_2}.

\begin{figure*}
\begin{center}
\includegraphics[width=0.6\textwidth]{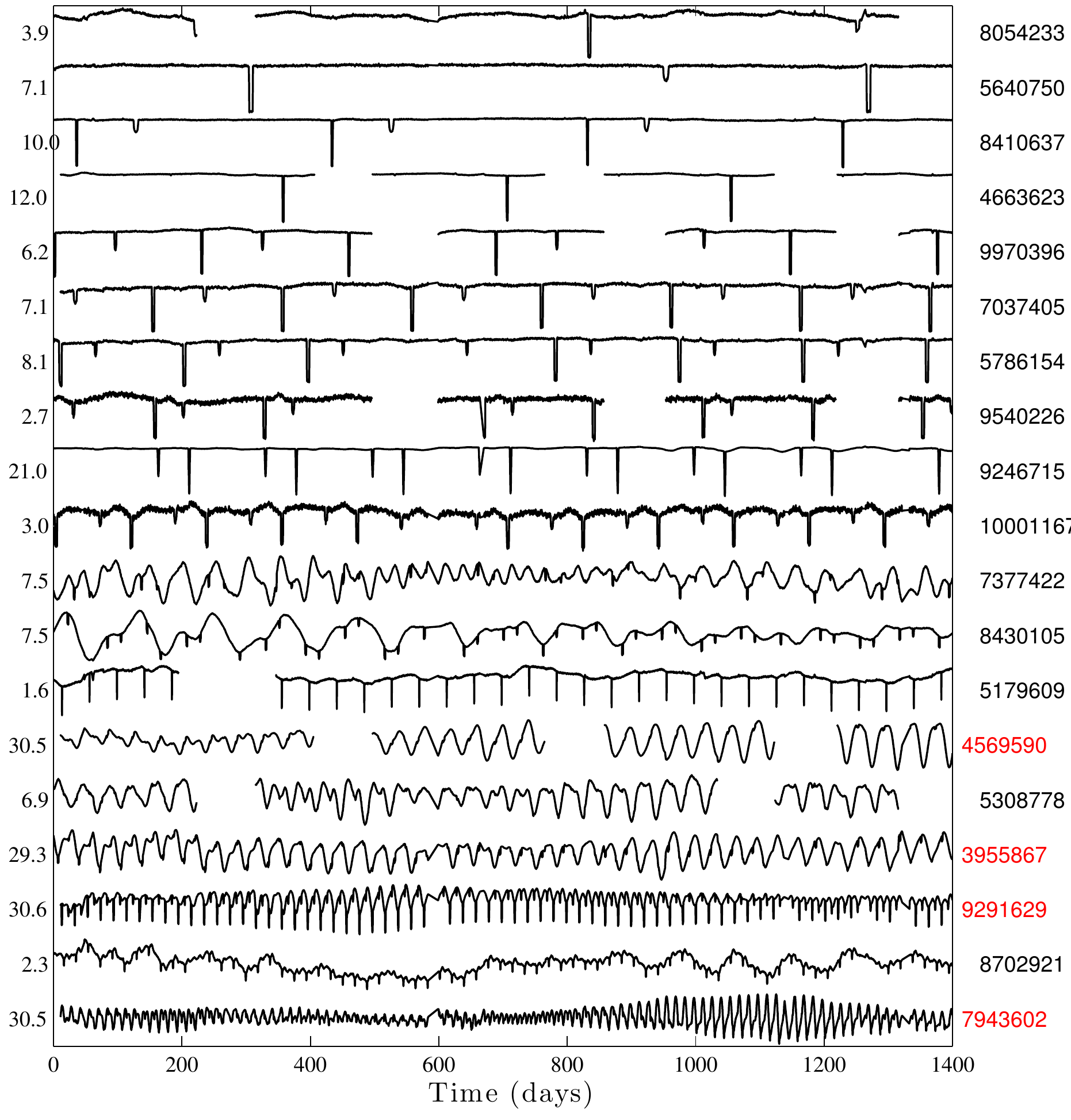}
  \caption{Full light curves of the 15 RGs in which solar-like oscillations are detected and the 4 where they are not detected, sorted by decreasing orbital period from top to bottom. Each system is labeled by its \textit{Kepler} ID number. This label is shifted to left and printed in red (online version) for those where no modes are detected. The $y$-axis indicates  the maximum variation in the flux (including eclipses), in \%. \label{fig_LC_var}}
\end{center}
\end{figure*}

\subsection{Two new candidates}
\label{sect_candidates}
From \citet{Slawson_2011}, EBs represent $\sim 1.4$\,\% of all targets observed by \textit{Kepler}. As presented in \citet{Gaulme_2013}, only 17 systems composed of at least one RG have been identified by cross-correlating both RG and EB catalogs. Even though EBs with RGs are likely less frequent than main-sequence binaries because orbital periods are necessarily longer, one must assume there are more EBs in the RG catalog that have not yet been detected.  Considering that the \citet{Slawson_2011} catalog is based on quarters Q0 to Q2 only, and that 16 quarters are now available, we did a systematic search for more EBs in the whole 14\,000 RG sample.

All time series were smoothed over 100 points (a bit more than 2 days) with a boxcar function, and the resulting time series were then differentiated for detecting possible eclipses.  For each dataset, an estimate of the root mean square (rms) noise level was first obtained by removing the outliers using the Peirce criterion \citep{Peirce_1852}. After removing the outliers, the start and end of eclipses were detected by using a 10-$\sigma$ threshold from the aforementioned rms noise level.  An eclipse is identified when both a negative and a positive value are detected within 10 days of each other. Then a possible following eclipse is determined by detecting the next flagged value that would occur later than 5 times the duration of the previous eclipse.  All parameters - smoothing factor of the time series, threshold, window for detection, etc.  -  can be adjusted depending on the star and the eclipse frequency (post facto). This method is very robust as derived from our initial experiments.  The detection technique was applied to each available quarter of the RG catalog.  The deepest detected eclipse on a single pass corresponds to a drop of 0.1\,\%.  Therefore the detection limit of the technique is about 0.014\,\% at $1\sigma$.

We detected 23 EB candidates that were not referenced in the literature, and among them, only three correspond to systems including at least one RG. The 20 others are false positives, either RGs aligned with EBs, or triple systems where a RG component is on a distant orbit with a main-sequence EB \citep{Gaulme_2013}. KIC 3458715 appears to be a possible heartbeat system, while KICs 4663623 and 8054233 appear to be \textit{bona fide} candidate RGEBs. In all three light curves, we detect clear solar-like oscillations typical of RGs. In this paper, we do not consider heartbeat systems as we base part of our analysis on the signal features occurring during eclipses. However, we report this possible new detection where the RG oscillations are characterized by $\nu\ind{max} = 170.5\pm0.7\ \mu$Hz and $\Delta\nu = 13.28\pm0.01\ \mu$Hz, corresponding to a radius $R = 5.21\pm0.03\ R_\odot$ and mass $M = 1.40\pm0.02\ M_\odot$. 

The system \textbf{KIC 4663623} has an orbital period of 358.08~day and displays the second-highest contrast between primary and secondary eclipse among all RGEBs we have identified so far.  The primary eclipse is 0.35\,\% deep whereas the secondary's depth is 10.9\,\%, which means that the companion star is significantly hotter than the RG. Another peculiarity with respect to other RGEBs of period longer than 200 days is its almost circular orbit ($e=0.04$). The asteroseismic mass and radius of the RG are $R_1=10.34\pm0.22\ R_\odot$ and $M_1=1.68\pm0.10\ M_\odot$, and the oscillation spectrum indicates that the RG is on the RGB. Also, it is a system with grazing eclipses, which makes it very challenging to model. Hence, the fitted parameters suffer from high uncertainty, with $R_2 = 0.8\pm0.1\ R_\odot$, $M_2=0.7\pm0.3\ M_\odot$, and $T_2\simeq10800$~K. The nature of the companion star is thus difficult to interpret and only radial velocities and atmospheric parameters from high-resolution spectroscopy will provide more constraints.

\textbf{KIC 8054233} has the longest orbital period of the RGEB sample, $P=1058.23$~day, on a rather eccentric orbit  $e=0.22$. The asteroseismic mass and radius of the RG are $R_1=10.86\pm0.21\ R_\odot$ and $M_1=1.65\pm0.09\ M_\odot$. The clear detection of mixed modes suggests that the RG belongs to the RC. As with most of the long-period systems from our sample, the companion star is likely an F-star, with radius $R_2=1.1\pm0.1\ R_\odot$, mass $M_2=2.0\pm0.3\ M_\odot$, and temperature $T_2\simeq6300$~K.  

\subsection{Noticeable changes for the previously known RGEBs}
The data analysis performed in \citet{Gaulme_2013} was based on \textit{Kepler} data through quarter Q11. Analysis through Q16 brings out significant changes for the following systems:

\textbf{KIC 4663185} was interpreted as a possible RG with a $\delta$-Scuti companion. It is actually a false positive resulting from blending from a nearby  $\delta$-Scuti star on \textit{Kepler}'s detector. We have removed it from the RGEB list.

\textbf{KIC 7377422} has a 107.62~day period, 0.44 eccentricity, and was classified among the systems with no oscillations. We now detect  RG oscillations of quite low amplitude and noisy structure with respect to most of the RGEBs (see Fig.~\ref{fig_PDS}). Nevertheless, its asteroseismic parameters can be  measured, giving radius and mass to be $R_1=9.42\pm0.84\ R_\odot$ and $M_1=1.03\pm0.28\ M_\odot$. The nature of its companion is consistent  with a solar analog with  $R_2=1.1\pm0.1\ R_\odot$,  $M_2=1.8\pm1.1\ M_\odot$, and $T_2 \simeq 5800$~K. 

\textbf{KIC 5640750} was the only system with unknown orbital parameters. Now its period is measured to be 987.40 days and eccentricity 0.32. This allows us to model the eclipses and infer the companion's nature, which appears to be  a subgiant F star as its radius, mass and temperature are $R_2=1.9\pm0.1\ R_\odot$, $M_2=1.7\pm0.3\ M_\odot$, and $T_2 \simeq 6600$~K. 

\textbf{KIC 5179609} was the only  RGEB with $\nu\ind{max}$ larger than the Nyquist frequency, and where no secondary eclipse was detected. Because of the extended dataset, the alias structure of oscillation modes is now well identified and leads to an accurate measurement of the asteroseismic parameters. The RG is the smallest of all RGEBs as its radius and mass are $R_1=3.67\pm0.08\ R_\odot$ and $M_1=1.36\pm0.09\ M_\odot$. We are now able to detect secondary eclipses in the folded light curve. However, it is very shallow (200 ppm), is not halfway between two consecutive primary eclipses, and displays a pure ``V''-shape. This means that this system has a rather eccentric orbit, and a secondary eclipse so grazing that the companion star is never fully eclipsed. The light curve modeling is thus extremely challenging, yet allowed us to constrain the eccentricity $e = 0.14\pm0.02$ and the companion size $R_2=0.4\pm0.1\ R_\odot$. However, mass and temperature estimates are not reliable. We believe that the companion is likely an M star.

\begin{table*}[h]
\begin{center}
\footnotesize
\renewcommand{\arraystretch}{1.} 
\caption{\small \it Relevant properties of the 19 systems from the KIC and from eclipse modeling. For each system,  modeling was performed on light curves from quarters 0 to 16, when applicable. The last two columns present the periods of stellar variability, when clearly identified, and the maximum peak-to-peak amplitudes measured on the whole time series. For the systems associated with a dagger symbol light-curve modeling is affected by  rather large uncertainties (see Sect. \ref{sect_candidates}).}
\begin{tabular}{l c c c c r c c c c c  c c}
\hline
\multicolumn{5}{c|}{KIC parameters}
&
\multicolumn{6}{|c}{LC modeling} &\multicolumn{2}{|c}{LC variability} \\
\hline
KIC           & $K\ind{mag}$ & $T\ind{eff}$ & $\log g$ & [Fe/H] & $P\ind{orb}$ & $e$ & $i$  &  $\displaystyle\frac{R_1 + R_2}{a}$&$\displaystyle\frac{R_2}{R_1}$&$\displaystyle\frac{T_2}{T_1}$  &$P\ind{var}$& $\displaystyle\left(\frac{\delta I}{I}\right)\ind{var}$  \\
                 &            & [K]        & [dex]   &  [dex] & [days] &     & [$^\circ$]    & [\%] &  & &[days] &[\%]\\
\hline
 8054233 & 11.783 &   4889 &   2.59 &  -0.36 & 1058.23 &   0.22 &  89.55 &   1.78 &   0.11 &   1.30 & \nodata &    0.5 \\
 5640750 & 11.565 &   4740 &   2.53 &  -0.46 & 987.40 &   0.32 &  89.99 &   2.65 &   0.13 &   1.40 & \nodata &    0.4 \\
 8410637 & 10.771 &   4873 &   2.76 &  -0.34 & 408.32 &   0.69 &  89.60 &   3.73 &   0.15 &   1.45 & \nodata &    0.2 \\
 4663623$^\dagger$  & 12.830 &   4980 &   3.30 &  -0.40 & 358.08 &   0.04 &  88.20 &   3.90 &   0.07 &   2.17 & \nodata &    0.5 \\
 9970396 & 11.447 &   4894 &   2.97 &  -0.26 & 235.30 &   0.20 &  89.43 &   4.42 &   0.14 &   1.28 & \nodata &    0.2 \\
 7037405 & 11.875 &   4778 &   2.68 &  -0.40 & 207.11 &   0.23 &  88.60 &   8.08 &   0.13 &   1.39 & \nodata &    0.3 \\
 5786154 & 13.534 &   4939 &   3.07 &  -0.26 & 197.92 &   0.37 &  89.12 &   7.05 &   0.15 &   1.37 & \nodata &    0.2 \\
 9540226 & 11.672 &   4758 &   2.35 &  -0.44 & 175.46 &   0.39 &  89.99 &   8.02 &   0.08 &   1.36 & \nodata &    0.3 \\
 9246715 &  9.266 &   4857 &   2.42 &  -0.40 & 171.28 &   0.35 &  87.11 &   7.80 &   0.84 &   1.04 &  93.30 &    2.0 \\
10001167 & 10.050 &   4845 &   2.53 &  -0.50 & 120.39 &   0.16 &  87.55 &  11.38 &   0.08 &   1.30 & \nodata  &    0.4 \\
 7377422 & 13.562 &   4683 &   2.62 &  -0.26 & 107.62 &   0.44 &  87.05 &   7.75 &   0.11 &   1.24 &  54.60 &    7.5 \\
 8430105 & 10.420 &   5143 &   2.78 &  -0.60 &  63.33 &   0.26 &  89.71 &   9.62 &   0.10 &   1.14 & 121.60 &    7.3 \\
 5179609$^\dagger$  & 12.776 &   4977 &   3.12 &   0.14 &  43.93 &   0.15 &  86.48 &   6.90 &   0.11 &   1.11 & 181.70 &    0.4 \\
 4569590 & 12.799 &   4764 &   2.78 &  -0.18 &  41.37 &   0.03 &  88.02 &  22.28 &   0.07 &   1.32 &  41.10 &   26.6 \\
 5308778 & 11.777 &   4963 &   2.73 &  -0.22 &  40.57 &   0.00 &  82.42 &  17.72 &   0.06 &   0.86 &  38.90 &    6.3 \\
 3955867 & 13.547 &   4880 &   3.26 &  -0.10 &  33.66 &   0.02 &  87.52 &  16.16 &   0.11 &   1.29 &  33.30 &   22.7 \\
 9291629 & 13.957 &   4809 &   3.11 &   0.08 &  20.69 &   0.01 &  84.06 &  23.70 &   0.23 &   1.28 &  20.60 &   16.4 \\
 8702921 & 11.980 &   5021 &   2.86 &   0.21 &  19.38 &   0.10 &  89.98 &  14.53 &   0.10 &   0.54 &  97.80 &    1.4 \\
 7943602 & 13.988 &   5096 &   3.40 &  -0.56 &  14.69 &   0.04 &  84.39 &  22.48 &   0.17 &   1.21 &  14.60 &   27.2 \\\hline
\end{tabular}
\label{tab_1}
\end{center}
\end{table*}


\begin{table*}[h]
\begin{center}
\footnotesize
\renewcommand{\arraystretch}{1.} 
\caption{\small \it Global seismic parameters and combined outputs from asteroseismic scaling relationships and light-curve modeling. (From left to right: ) $\nu\ind{max}$ is the frequency at maximum amplitude of the oscillation mode envelope, $\Delta\nu$ the modes' observed large separation, $A\ind{max}$ the amplitude of the largest mode.  $R_1$ and $M_1$ the radius and mass of the RG component, $R_2$, $M_2$, and $T_2$ the radius, mass, and effective temperature of the companion star. Class - RG refers to the RG class (RGB/RC/RC2/AGB), and Class - Cmp to the possible spectral type of the companion star. A star symbol associated with the companion's class indicates that the classification is uncertain, and a question mark indicates that no classification could be drawn. }
\begin{tabular}{l c c c c c c c l l l  }
\hline
&\multicolumn{3}{|c|}{Mode properties}
&
\multicolumn{2}{|c}{Asteroseismic scaling} &\multicolumn{3}{|c}{Combined outputs} &\multicolumn{2}{|c}{Class} \\
\hline
KIC           & $\nu\ind{max}$ & $\Delta\nu$ & $A\ind{max}$ & $R_1$         & $M_1$         & $R_2$        & $M_2$        & $T_2$ & RG & Cmp\\
                 &   [$\mu$Hz]         &   [$\mu$Hz] & [ppm]               & [$R_\odot]$ &[$M_\odot]$ &[$R_\odot]$ &[$M_\odot]$&   [K]     &        &         \\
\hline
  8054233 & $47.32 \pm 0.29$ & $4.806 \pm 0.005$ &  16.6 & $10.86 \pm 0.21$ & $ 1.65 \pm 0.09$ & $  1.1 \pm 0.1$ & $  2.0 \pm 0.3$ &  6342 & RC & F \\
  5640750 & $24.40 \pm 0.16$ & $2.987 \pm 0.005$ &  47.3 & $14.27 \pm 0.31$ & $ 1.45 \pm 0.09$ & $  1.9 \pm 0.1$ & $  1.7 \pm 0.3$ &  6618 &RGB/AGB  & F$^\star$\\
  8410637 & $44.80 \pm 0.30$ & $4.640 \pm 0.010$ &  20.5 & $11.01 \pm 0.26$ & $ 1.61 \pm 0.11$ & $  1.6 \pm 0.1$ & $  1.5 \pm 0.3$ &  7042 &  RC& F\\
  4663623 & $52.38 \pm 0.39$ & $5.205 \pm 0.005$ &  21.4 & $10.34 \pm 0.22$ & $ 1.68 \pm 0.10$ & $  0.8 \pm 0.1$ & $  0.7 \pm 0.3$ & 10785 & RGB  & ? \\
  9970396 & $63.19 \pm 0.23$ & $6.336 \pm 0.005$ &  19.7 & $ 8.35 \pm 0.15$ & $ 1.31 \pm 0.07$ & $  1.2 \pm 0.1$ & $  1.1 \pm 0.2$ &  6284 &  RGB& F \\
  7037405 & $21.93 \pm 0.16$ & $2.765 \pm 0.005$ &  48.9 & $15.03 \pm 0.35$ & $ 1.45 \pm 0.10$ & $  1.9 \pm 0.1$ & $  1.4 \pm 0.3$ &  6632 & RGB& F$^\star$\\
  5786154 & $30.06 \pm 0.17$ & $3.518 \pm 0.005$ &  54.0 & $12.94 \pm 0.27$ & $ 1.50 \pm 0.09$ & $  1.9 \pm 0.1$ & $  1.7 \pm 0.3$ &  6771 & RGB& F$^\star$\\
  9540226 & $27.74 \pm 0.17$ & $3.217 \pm 0.005$ &  31.1 & $14.01 \pm 0.26$ & $ 1.59 \pm 0.08$ & $  1.2 \pm 0.1$ & $  1.4 \pm 0.2$ &  6450 &  RGB& F\\
 9246715 & $106.38 \pm 0.75$ & $8.327 \pm 0.010$ &   6.6 & $ 8.10 \pm 0.18$ & $ 2.06 \pm 0.13$ & $  6.8 \pm 0.1$ & $  1.1 \pm 0.3$ &  5060 &RC2& RG\\
 10001167 & $19.99 \pm 0.14$ & $2.759 \pm 0.005$ &  54.5 & $13.85 \pm 0.32$ & $ 1.13 \pm 0.07$ & $  1.1 \pm 0.1$ & $  1.0 \pm 0.2$ &  6321 & RGB& F\\
  7377422 & $40.12 \pm 3.01$ & $4.701 \pm 0.005$ &   9.2 & $ 9.42 \pm 0.84$ & $ 1.03 \pm 0.28$ & $  1.1 \pm 0.1$ & $  1.8 \pm 1.1$ &  5823 &  ? & G \\
  8430105 & $66.80 \pm 1.64$ & $7.130 \pm 0.010$ &   5.4 & $ 7.14 \pm 0.28$ & $ 1.04 \pm 0.12$ & $  0.7 \pm 0.1$ & $  0.8 \pm 0.3$ &  5846 &  RGB & K\\
  5179609 & $336.87 \pm 2.20$ & $22.160 \pm 0.030$ &   4.1 & $ 3.67 \pm 0.08$ & $ 1.36 \pm 0.09$ & $  0.4 \pm 0.1$ & $  0.1 _{-0.1}^{+0.2}$ &  5508 &  RGB & ?\\
  4569590 &           \nodata         &     \nodata                 &\nodata&            \nodata       &\nodata&            \nodata         &\nodata&   6311 & \nodata & F \\
  5308778 & $44.88 \pm 2.05$ & $5.120 \pm 0.050$ &   4.3 & $ 9.14 \pm 0.72$ & $ 1.12 \pm 0.24$ & $  0.6 \pm 0.1$ & $  0.2_{-0.2}^{+0.6}$ &  4249 & ?  &M$^\star$ \\
  3955867 &         \nodata           &     \nodata                 &\nodata&          \nodata        &\nodata&          \nodata        &\nodata&   6281 & \nodata  & F \\
  9291629 &         \nodata           &     \nodata                 &\nodata&          \nodata        &\nodata&          \nodata        &\nodata&   6179 & \nodata   & F \\
  8702921 & $192.70 \pm 0.80$ & $14.070 \pm 0.020$ &   3.2 & $ 5.23 \pm 0.11$ & $ 1.58 \pm 0.09$ & $  0.5 \pm 0.1$ & $  0.6 \pm 0.2$ &  2727 & RGB  & M\\
  7943602 &         \nodata           &     \nodata                 &\nodata&          \nodata        &\nodata&          \nodata        &  &\nodata &  \nodata  & F\\ 
\hline
\end{tabular}
\label{tab_2}
\end{center}
\end{table*}

\section{Binarity, activity, and oscillation amplitudes}
\label{sect_obs}

\subsection{Red-Giant Activity}
An important feature  observed in the light curves is the presence of quasi-periodic modulations for the nine systems with orbital periods shorter than 110 days. The entire light curves for all 19 RGEBs are shown in Fig.~\ref{fig_LC_var}. The modulations in the short-period binaries display low amplitudes for KICs 5179609 and 8702921, 0.4 and 1.4\,\% respectively, and high amplitude for the others, ranging from 6 to 27\,\% (as listed in Table \ref{tab_1}). For systems with longer periods, fluctuations are less than 0.5\,\%, except for the double RG KIC 9246715 where it reaches 2\,\% (see Sec.~\ref{sect_9246715}). Such photometric fluctuations are typical of surface activity commonly detected on main-sequence stars observed by CoRoT and \textit{Kepler} \citep[e.g.][]{Mosser_Baudin_2009,Bonomo_Lanza_2012}. 
Consider that groups of starspots are present on one of the two stars in the binary system. Given that most of the systems are likely composed of an RG and a much smaller star  ($R_2/R_1 \leq0.23$, except KIC 9246715), the companion represents at most about 10\% of the total luminosity, which makes it  very unlikely that spots belong to the smaller star.

\begin{figure}
\includegraphics[width=0.47\textwidth]{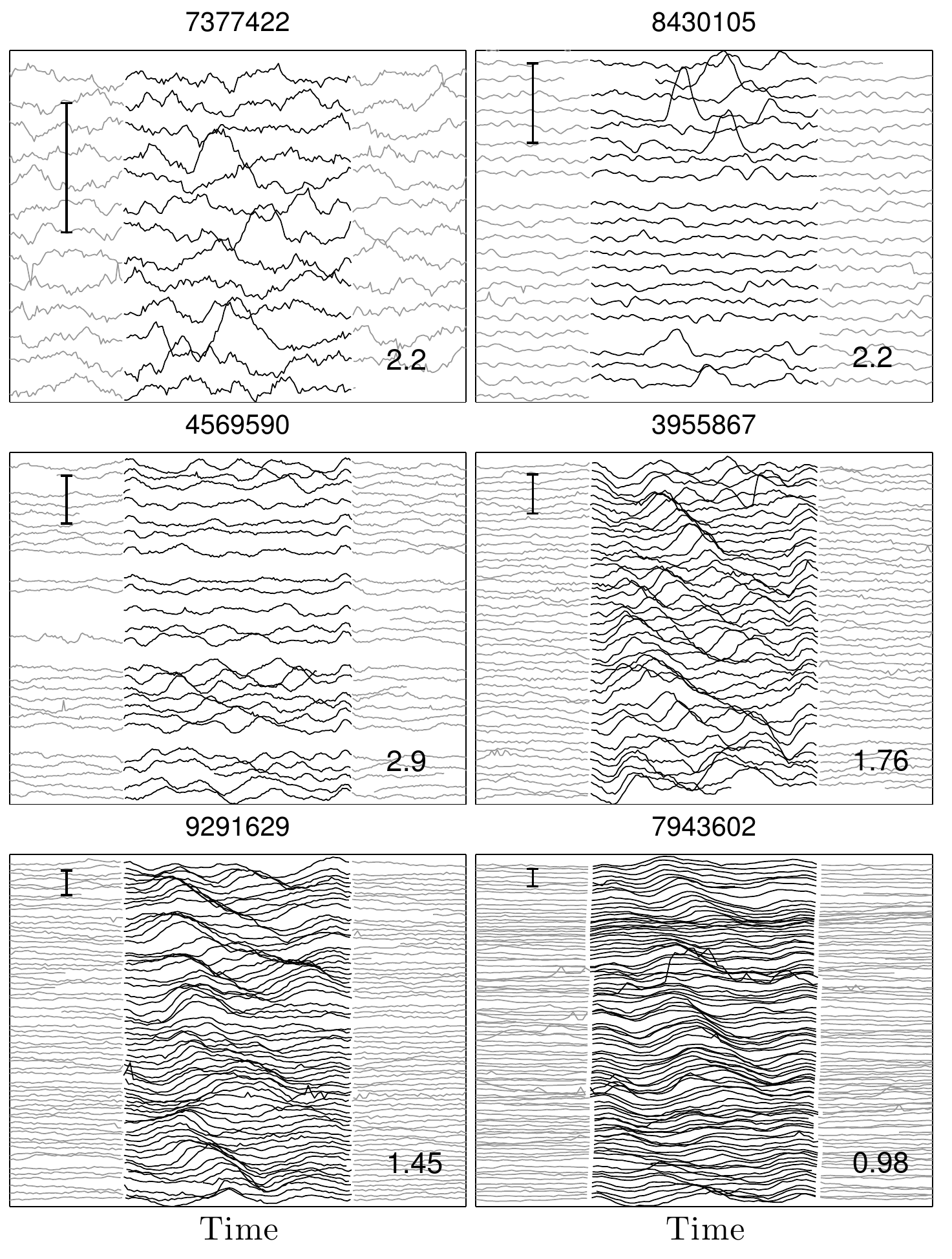}
\caption{Extracts of light curves during the primary eclipses (companion eclipses the  RG) for six of the nine systems where high variability is detected. The black regions correspond to the eclipses, while the gray regions are out of eclipse. Each row corresponds to one eclipse, and they are vertically sorted by date (top: first, bottom: last).  The vertical line on the top left of each sub-panel indicates photometric fluctuations of 1\,\%.  The time axis is normalized to the primary eclipse duration, indicated in days in the bottom right of each sub-panel. \label{fig_RG_spots}}
\end{figure}

In addition, upon dividing the original light curves by the corresponding eclipse models, six of the nine short-period systems display significant photometric fluctuations during primary eclipses (illustrated in Fig.~\ref{fig_RG_spots}). This phenomenon mostly increases as the orbital period decreases. We interpret this as the companion stars alternatively passing in front of darker and brighter regions on the RG surfaces, as has been observed for exoplanets transiting main-sequence stars \citep[e.g.][]{Silva-Valio_2010}. 
For these systems, we performed spot modeling  as described in \citet{Mosser_Baudin_2009}. Even though it cannot provide a precise measurement of stellar inclination, models with inclinations close to 90$^\circ$ provide the best solutions. In other words, spot modeling shows no sign of spin-orbit misalignment in our systems, thus suggesting evidence of differential rotation. The resemblance of the primary-eclipse light curves in three cases where the orbital and variability periods are equal and no oscillations are detected indicates that these stars follow a similar differential rotation profile. 
The dedicated spot analysis will be done in a forthcoming paper. We note here in the eclipse light curve of KICs 3955867, 9291629 and 7943602 (Fig.~\ref{fig_RG_spots}) a small delay of the spot structures with respect to the eclipses, with a relative difference of about 0.23, 1.05 and 0.25\,\% respectively per orbital period. This is the signature of differential rotation. If we assume the equatorial rotation period to be equal to the orbital rotation period, this implies either that the main spot structures have low latitudes, or that due to binarity the differential factor is significantly smaller than observed in single stars.

\begin{figure}
\begin{center}
\includegraphics[width=0.47\textwidth]{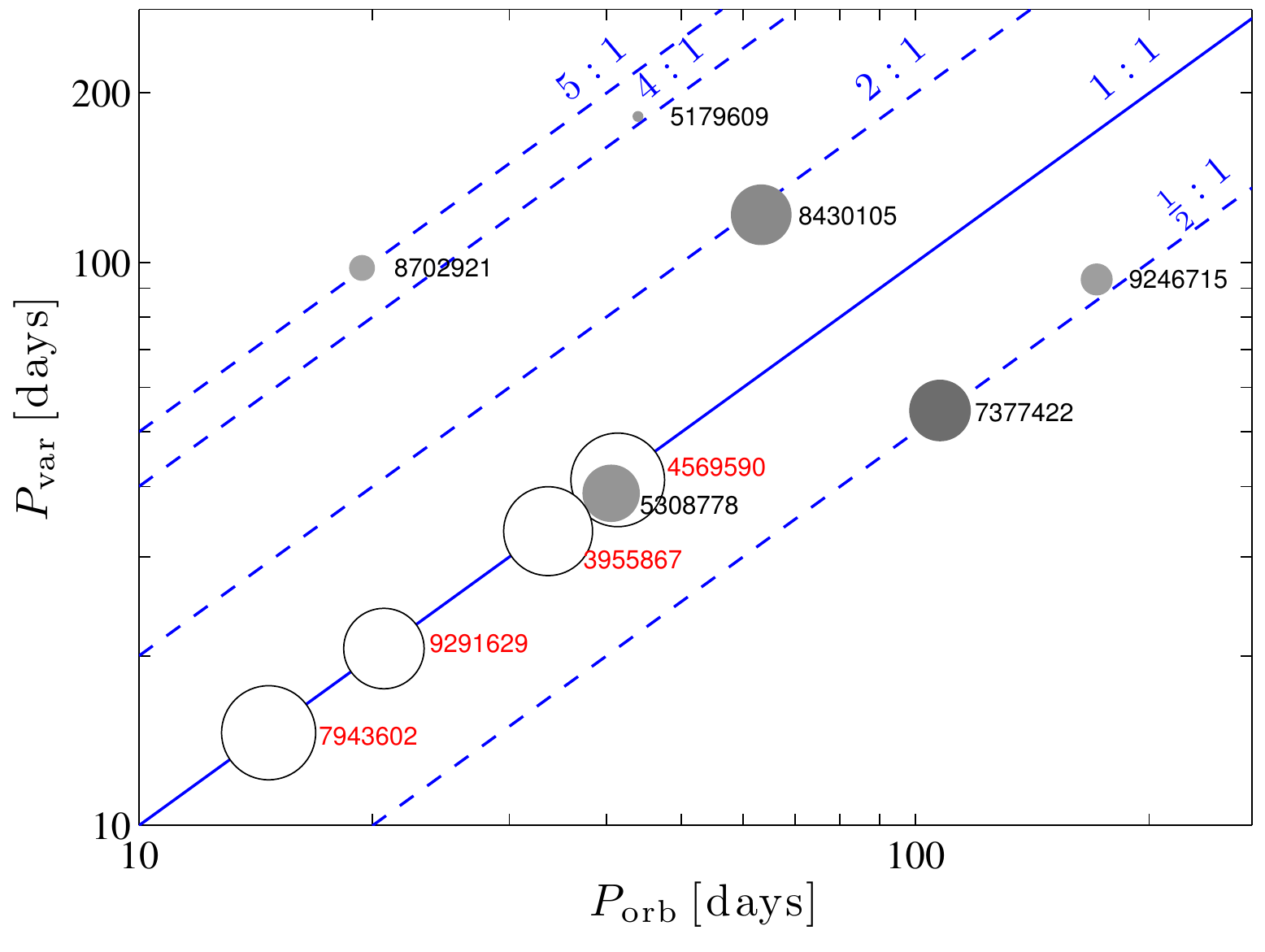}
\caption{Orbital period vs.~variability period for the 10 systems where significant stellar activity is detected. The size of each symbol represents the amplitude of stellar variability, and the gray scale indicates the pulsation-mode amplitude (white - no modes; black - large amp). The parallel lines (plotted in blue in online version) indicate isolevels of the ratio $P\ind{var}/P\ind{orb}$.  \label{fig_resonance}}
\end{center}
\end{figure}

Beyond the interest that RG variability represents, it carries precious information about binary systems, in particular on the RG surface rotation. We measured the fundamental  periods of variability $P\ind{var}$ (see Appendix II) of the ten light curves where activity is detected (including KIC 9246715), and we plot them as a function of the orbital periods in Fig.~\ref{fig_resonance}. The five systems with the shortest orbits display almost equal variability and orbital periods. The remaining five systems, respectively, display variability periods of about one half (KICs 7377422 \& 9246715), twice (KIC 8430105), four times (KIC 5179609) and five times (KIC 8702921) their orbital periods. This demonstrates that stellar activity is related to tidal interactions as all variability periods are resonances of their orbits. 
The five systems with the shortest periods appear to be synchronized in phase, RG rotation + binary orbit, and circularized, as their eccentricities are practically zero. It is impossible to know about possible companion synchronization (rotation) as its photometric signature is extremely weak.

\subsection{Oscillation Amplitude}

\begin{figure}
\begin{center}
\includegraphics[width=0.45\textwidth]{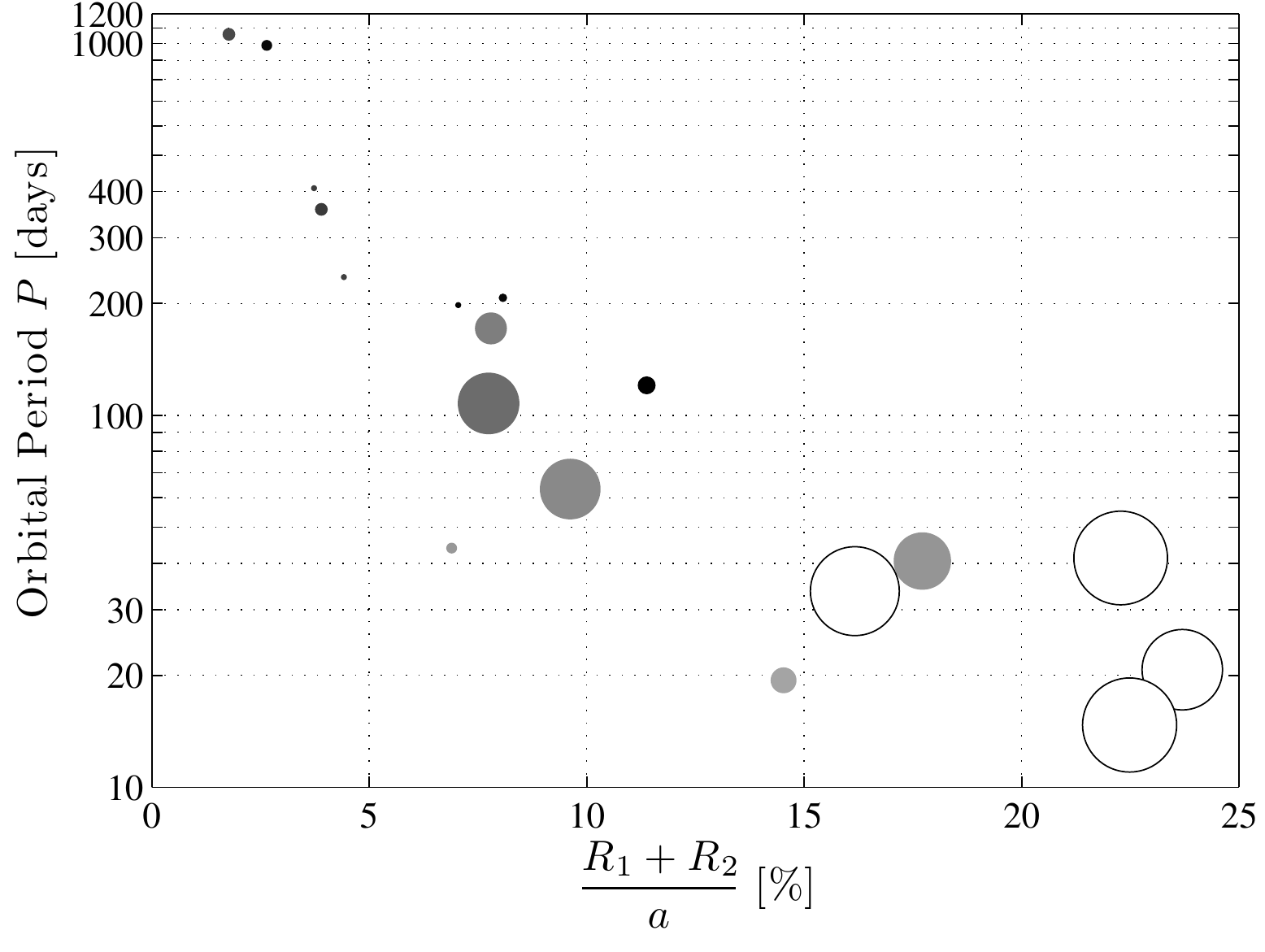}
\includegraphics[width=0.45\textwidth]{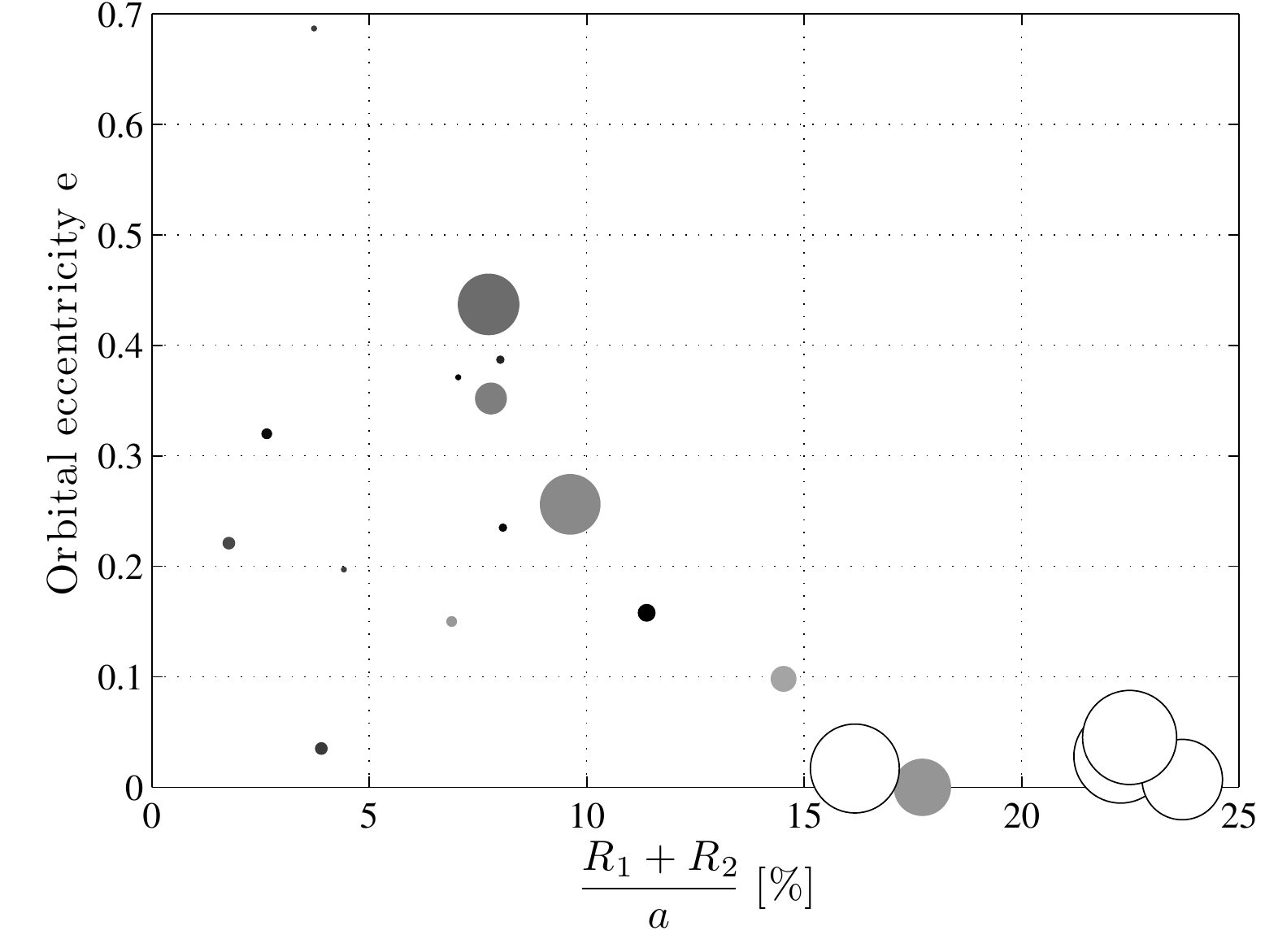}
\caption{(\textit{Top}:) Orbital period vs.~relative radii for the 19 systems. The size of each symbol represents the amplitude of stellar variability, and the gray scale indicates the pulsation mode amplitude (white - no modes; black - large amp.) (\textit{Bottom}:) The same but for the eccentricity of the binary system.  \label{fig_sumR_P_e}}
\end{center}
\end{figure}

The major connection regarding the absence of modes in four of the 19 RGEBs is that they are the systems with highest photometric variabilities (17 to 27\,\% peak-to-peak). They also constitute four of the five systems that are synchronized, i.e., with equal variability and orbital periods, four of the six systems with the shortest orbits, four of the six systems with the smallest eccentricities, and four of the five systems with the largest sum of relative radii (Fig. \ref{fig_sumR_P_e}). Regarding the latter, when this ratio is larger than 18\%, we never detect modes, whereas when it is lower than 16\%, we always detect oscillation modes.

\begin{figure}
\begin{center}
\includegraphics[width=0.45\textwidth]{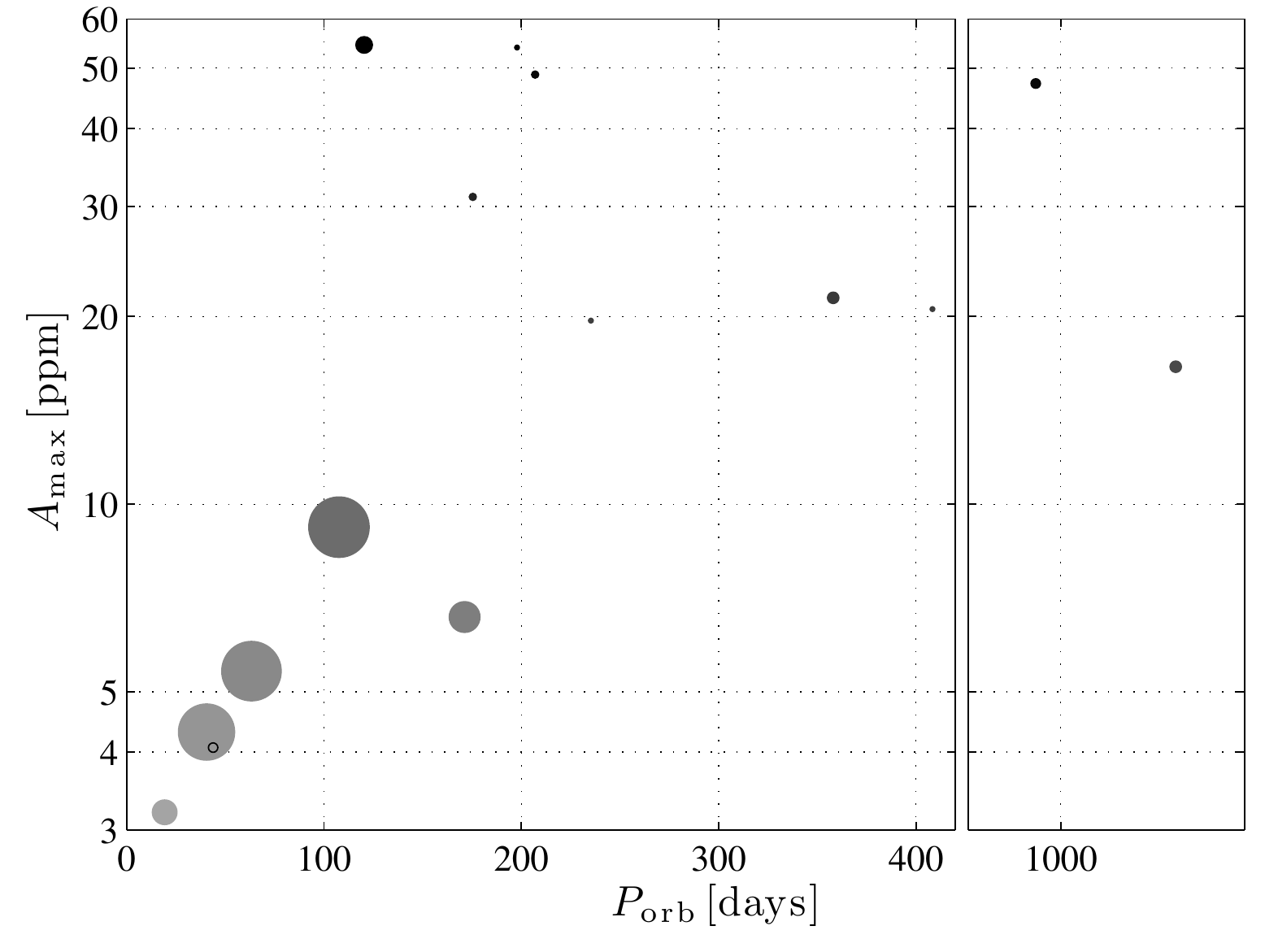}
\includegraphics[width=0.45\textwidth]{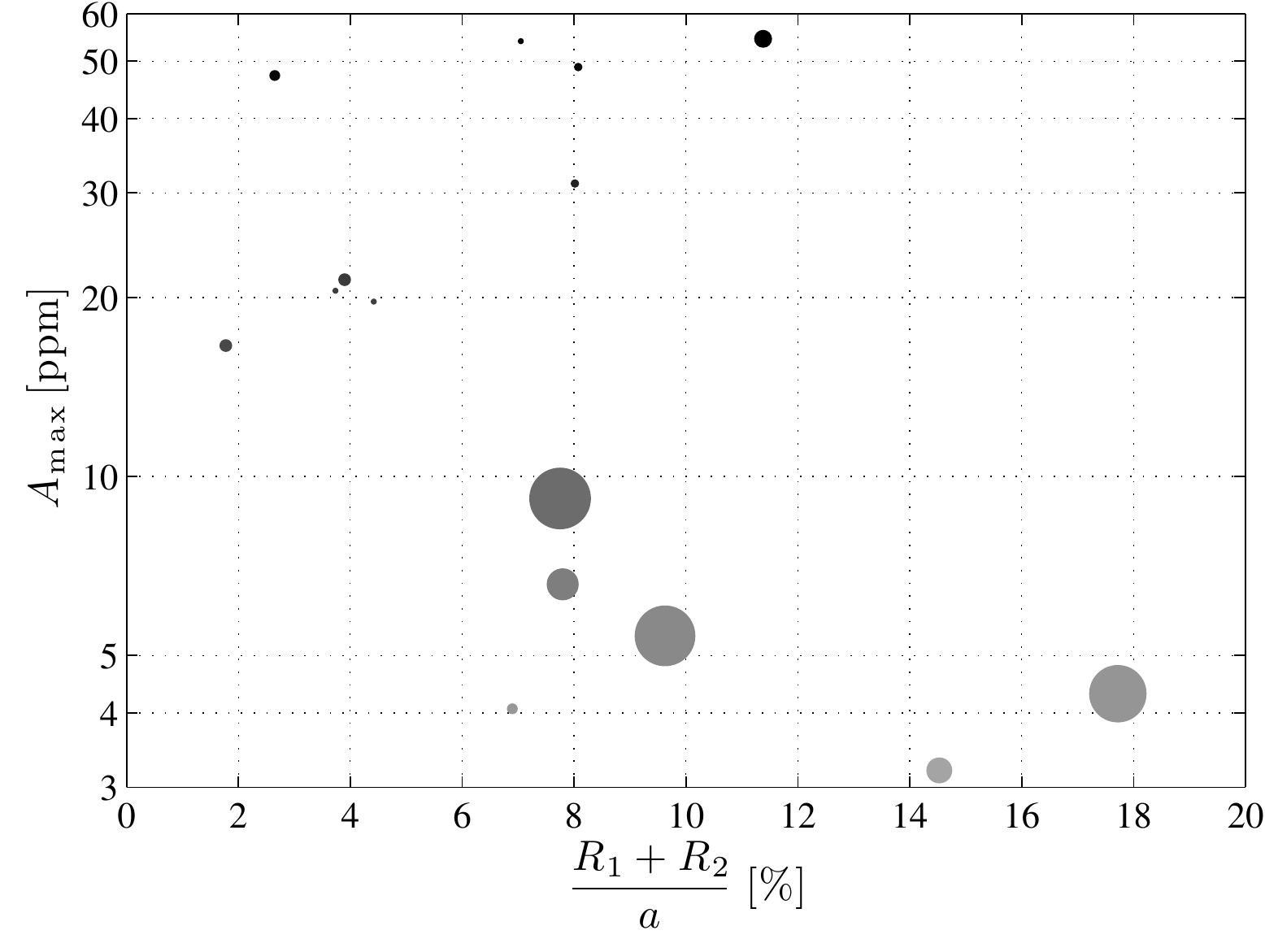}
\caption{(\textit{Top}:) Maximum mode amplitude vs.~orbital period for the 15 systems displaying solar-like oscillations. The size of each symbol represents the amplitude of stellar variability, and the gray scale indicates the mode amplitude (light gray - low amp.; black - large amp.). (\textit{Bottom}:) The same as above but with the sum of relative radii taking the place of orbital period. \label{fig_Amax_P}}
\end{center}
\end{figure}

An additional interesting observable is the oscillation mode amplitude of the five systems where significant variability as well as oscillations are detected (KICs 7377422, 8430105, 5179609, 5308778, 8702921, but 9246715). We obtain a proxy of mode amplitude by measuring the largest peak's height in the PSD after background subtraction. Then, the amplitude $A_{\rm max}$ - the square root of the latter - is normalized by the factor $(1 + R_2^2 T_2^4/R_1^2 T_1^4)$ to take into account the oscillation dilution related to the companion star. In Fig.~\ref{fig_Amax_P}, we plot the mode amplitudes as a function of orbital period. It shows that the mode amplitude is reduced for all systems displaying a significant variability, since their amplitudes range from  3.2 to 9.2 ppm. Moreover, the logarithm of the mode amplitude displays a linear increase as a function of orbital period for these five systems. For all other systems with no observed activity, the mode amplitude ranges from 16.6 to 54.5 ppm with no specific dependence on orbital period. No such dependence is observed as a function of the sum of relative radii (Fig.~\ref{fig_Amax_P}, bottom).

To explain this observational dependence, three possible biases may be considered for the mode-depleted systems. Firstly, small RGs may pulsate at higher frequencies than the Nyquist frequency associated with the 30-min cadence (shorter cadence data for most RGs are unavailable). However, as for KIC 5179609, which is the smallest RG in the RGEB sample, aliases of the modes should clearly be detected in the power spectra. Secondly, out of the seven faintest targets, four of them do not display oscillations ($12.8\leq m\ind{Kep} \leq 14.0$). However, KIC 5786154 has modes with the second-highest  amplitude, yet it is one of the faintest systems ($m\ind{Kep} = 13.5$),  at the same level of the non-pulsating KIC 3955867. Thirdly, if not properly filtered out, harmonics of stellar variability dominate the PSD up to the Nyquist frequency, because it is a complex signal far from a pure sinusoid. To efficiently remove the variability contribution, we filter the times series with a weighted moving average that erases most of the signal of frequencies lower than 20\,$\mu$Hz. Thus, modes could hide in this frequency range.  This is unlikely for the four close-in systems with $(R_1+R_2)/a \geq 16$\,\% and $P\leq 41$ days. In such systems, RGs with radii larger than $R\geq10\ R_\odot$ seem unphysical \citep{Gaulme_2013, Frandsen_2013}, and we know from asteroseismic scaling relations \citep[e.g.][]{Huber_2011} that a $10\ R_\odot$ RG displays its maximum amplitude modes between 30 and $70\ \mu$Hz, i.e., out of the filtered range.

\begin{figure}
\begin{center}
\includegraphics[width=0.49\textwidth]{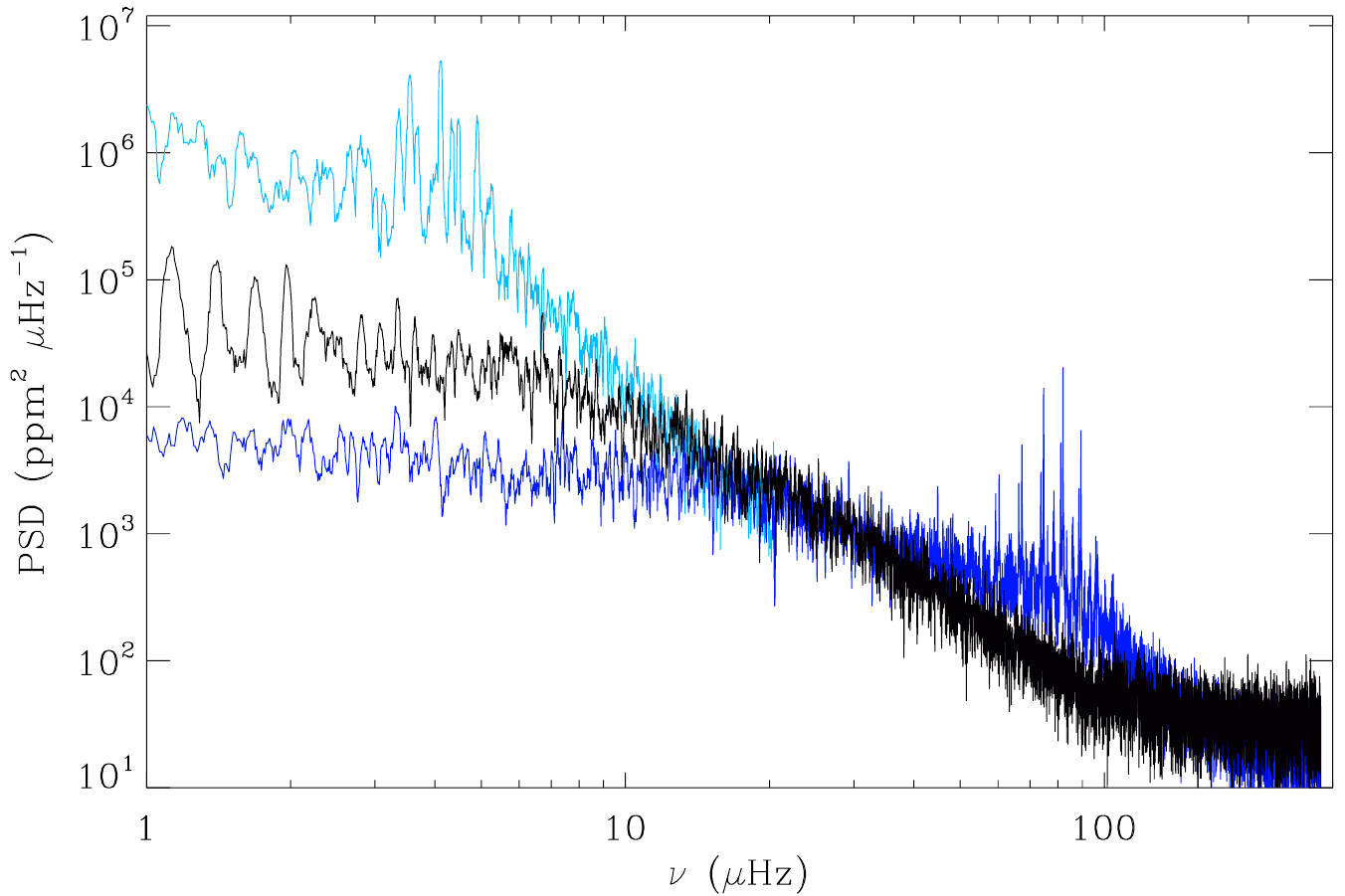}
\caption{Comparison between power spectra of the non-pulsating target KIC 4569590 (black line) with two RGs of similar magnitude and regular oscillation modes (plotted in blue in online version), whose $\nu\ind{max}$ are about 4 and 80 $\mu$Hz.\label{fig_PDS_comp}}
\end{center}
\end{figure}

In Fig.~\ref{fig_PDS_comp}, we plot the PSD of one of the four non-pulsating RGEBs together with the PSD of two non-EB RGs of equal photometric magnitudes observed with \textit{Kepler}. The result is similar for the other three systems with no oscillations and we do not plot their PSDs in Fig.~\ref{fig_PDS_comp}. It clearly shows that the power spectra are not significantly contaminated by harmonics of orbital periods, as the background levels are very close to those observed in the lonesome RGs. It also indicates that if modes of regular amplitude were present in the datasets, they would have shown up in the PSD. We conclude that the mode depletion in these four systems is real.

\subsection{The double RG with only one oscillator}
\label{sect_9246715}
The case of the double-RG system, KIC 9246715, stands apart from the rest and tends to confirm what we have described above. Its apparent magnitude of 9.3 is the brightest of the whole sample, making it the ideal candidate for identifying solar-like oscillations.  However, oscillations are detected for only one of the two RGs and their amplitudes are low ($A\ind{max} \simeq6.6$~ppm) with respect to the RGs with similar global seismic parameters. From asteroseismic scaling relations, the RG that displays pulsations has a radius $R_1 = 8.10\pm0.18~R_\odot$ and a mass $M_1 = 2.06\pm0.13~M_\odot$. By combining this with the output of eclipse modeling, it turns out that the companion's radius and mass are either $(6.8\pm0.1~R_\odot, 1.1\pm0.3~M_\odot)$ or $(9.7\pm0.2~R_\odot, 3.3\pm0.5~M_\odot)$. The latter solution looks unlikely in terms of the companion's mass, so we suspect that oscillations are associated with the larger of the two stars. A detailed study of this system, including radial-velocity analysis, is in progress and should confirm this.

\begin{figure}
\includegraphics[width=0.45\textwidth]{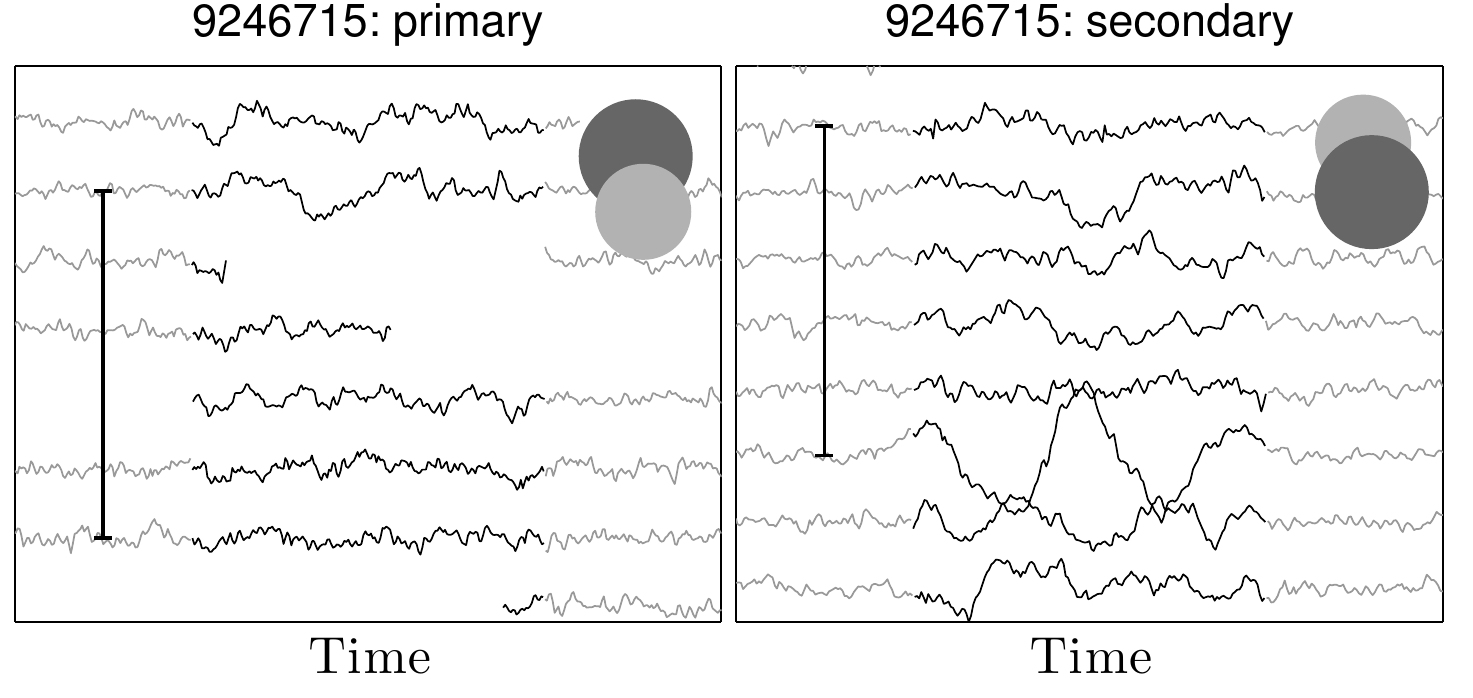}
\caption{Extracts of light curves during the primary (left) and secondary eclipses (right) for the double RG 9246715. The black regions correspond to the eclipses while the gray to out of eclipses.  The vertical line on top left of each sub-pannel indicate photometric fluctuations of 1\,\%. Eclipse durations are respectively 3.2 and 3.8 days for primary and secondary eclipses. The gray disks represent the eclipse configuration, where light gray indicates higher surface temperature than dark gray. \label{fig_9246715} }
\end{figure}

We detect photometric modulations of 2\,\% amplitude, which is quite large with respect to systems with similar orbital periods. Spot signatures are clearly detected during the deepest eclipses, with amplitudes ranging from 1000 to 4000~ppm peak-to-peak, whereas spots are at the detectability limit during the shallowest eclipses, with amplitudes reaching at most 1400~ppm (Fig.~\ref{fig_9246715}). The eclipse modeling indicates that the deepest eclipses correspond to the situation where the bigger of the two stars eclipses the smaller. Because of an inclination quite far from $90^\circ$ for a rather long-period system ($87.1^\circ$), a high eccentricity (0.35), and a low relative radii (7.80\,\%), eclipses are only partial and the bigger RG covers at maximum about 48\,\% of the smaller's diameter, while the smaller covers about 32\,\% of the bigger's (see configuration on Fig~\ref{fig_9246715}). It thus appears that spots are likely more present on the smaller of the RGs, where no solar-like oscillations are measured. Note that such an interpretation is not valid if both stellar rotation axes are perpendicular to the orbital plane and if spots are limited to the equatorial region (latitudes less than $\pm20^\circ$). Indeed, in such a case, a 32-\% eclipse magnitude means that the equator (latitude 0$^\circ$) of the eclipsed star is never hidden and only latitudes higher than about 20$^\circ$ are hidden on one hemisphere.

This example is particularly important because both stars are similar in terms of luminosity ($L_2/L_1 = 0.83$), and hence in terms of signal-to-noise ratio. It should therefore be possible to detect solar-like oscillations from each star from the light curve. The only significant difference between them, in terms of physical properties, is their masses, which implies asymmetric tidal forcing in the system. Moreover, it confirms that stellar activity, as starspots, can be present on RG stars. It also shows that the more active the RG, the weaker the apparent mode signature.

\section{Discussion and Conclusion}
\label{sect_discussion}

We have shown that the absence of detected solar-like oscillations in four of the 19 RGEBs is part of a bigger picture: the shorter the orbital period and the closer the stars are, the weaker the oscillation modes become, until complete mode depletion is reached. This is observed in the closest systems where rotational and orbital periods are almost synchronized and where strong surface activity is detected. Past works give context and help interpreting this original result.

 \citet{Chaplin_2011b} studied the impact of stellar surface activity on the detectability of solar-like oscillations for mostly main-sequence and sub-giant stars that were observed by \textit{Kepler}. They measured the light-curve variability, in a similar manner as presented here, for about 500 solar-like stars that were observed at short cadence. They found evidence of a clear depletion of detected modes for stars with increased level of activity, attributing this to the presence of magnetic fields. They pointed out that magnetic structures in the solar photosphere are known to absorb acoustic oscillations and that strong magnetic fields can diminish the turbulent velocities in a convectively unstable layer, altering the mechanism for the  driving of acoustic modes. However, \citet{Huber_2011} did not confirm this observation for the RGs of their sample.

For RGs, evidence of magnetic field and surface activity has been discussed  in the past, even though RGs were thought to be inactive due to decreased rotation during their late evolution. Single RG stars known as chromospherically active were identified by detection of Ca\,{\sc ii} H and K emission \citep[e.g.][and refs. therein]{Dupree_Smith_1995,Auriere_2011,Konstantinova_2012}. The Zeeman Doppler imaging technique \citep[][hereafter ZDI]{Donati_Brown_1997} allowed for mapping of the magnetic surface of such giants and revealed the presence of spots. In particular, the recent and preliminary work of \citet{Konstantinova_2013} reports the detection of magnetic fields in 15 out of a sample of 29 G, K, and M single giants observed with ZDI. Active RGs likely constitute a minority of cases. The origin of their magnetic field and activity is thought to be generated either through dynamo mechanisms resulting from planet engulfment \citep{Siess_Livio_1999}, angular dredge-up from the interior for the fastest rotators \citep{Simon_Drake_1989}, or from remnant rotation for massive stars \citep{Stepien_1993}.
Active RGs were also identified in rather close detached  binary systems that usually rotate fast, with orbits and rotations in phase or close to a resonance, and where spots are detected with ZDI \citep[][]{Ozdarcan_2010,Olah_2013}.


The case of the hierarchical triply eclipsing system HD~181068 \citep[KIC 5952403,][]{Derekas_2011}, composed of a RG in a 45-day orbit with a pair of red dwarfs in a close 0.9-day orbit, displays common properties with our systems. \citet{Fuller_2013} carefully analyzed 11 \textit{Kepler} quarters of this system and could not detect any solar-like oscillations of the RG, even though its mass, radius, effective temperatures, and hence its $\nu\ind{max}$, are known from eclipse modeling and radial velocities. Their analysis indicates that the RG is tidally synchronized with the 45-day double dwarf orbit, and they suggest that a high magnetic activity resulting from such a rapid rotation suppresses the excitation of solar-like oscillations.

For the systems studied here, we  suggest  that binarity spins up RGs, with this phenomenon becoming stronger as systems are closer. This leads to the development of a dynamo mechanism, and thus the generation of  magnetic fields in the RGs that become visible at the surface. The resulting spots likely absorb part of the pressure mode energy making oscillations impossible to detect in the closest systems. Alternatively, we may think that the presence of spots shows that the convective energy is diverted into activity signal and not into global oscillations. This would mean that the properties of convection are considerably affected by binarity in the closest systems, and that oscillation excitation is reduced, or suppressed altogether.
For the short-period systems KIC 8702921 and KIC 5308778, the orbits are almost circular, the rotational periods are harmonics of the orbital periods, and the orbital periods are typical of the other systems with no oscillations. We likely detect low-amplitude oscillations only because their companions are low-mass M-stars that generate rather weak tidal forces, as is also evident by the low amplitudes of their surface activity. Finally, this reasoning is consistent with the case of KIC 9246715, where the stellar masses are also a factor of two different and oscillations modes are detected only on the RG with larger mass. 

We have discovered two new \textit{bona fide} RGEBs and one heartbeat system candidate. KIC 8054233 is the longest-period RGEB in this work, and KIC 4663623 the system in which the temperature contrast  between the two components is the largest ($\gtrsim 2$). We also updated the \citet{Gaulme_2013} list by determining the orbital parameters of KIC 5640750, properly characterized the seismic parameters of KIC 5179609 whose $\nu\ind{max}$ is larger than the Nyquist frequency, detected oscillations in KIC 7377422, and removed a false-positive. Efforts in obtaining radial velocities of all systems and performing asteroseismic modeling on individual frequencies of RG oscillations are in progress in collaboration with the red-giant group of the Kepler asteroseismic science consortium.

\appendix

\section{Light-curve processing}
\begin{figure}
\begin{center}
\includegraphics[width=0.45\textwidth]{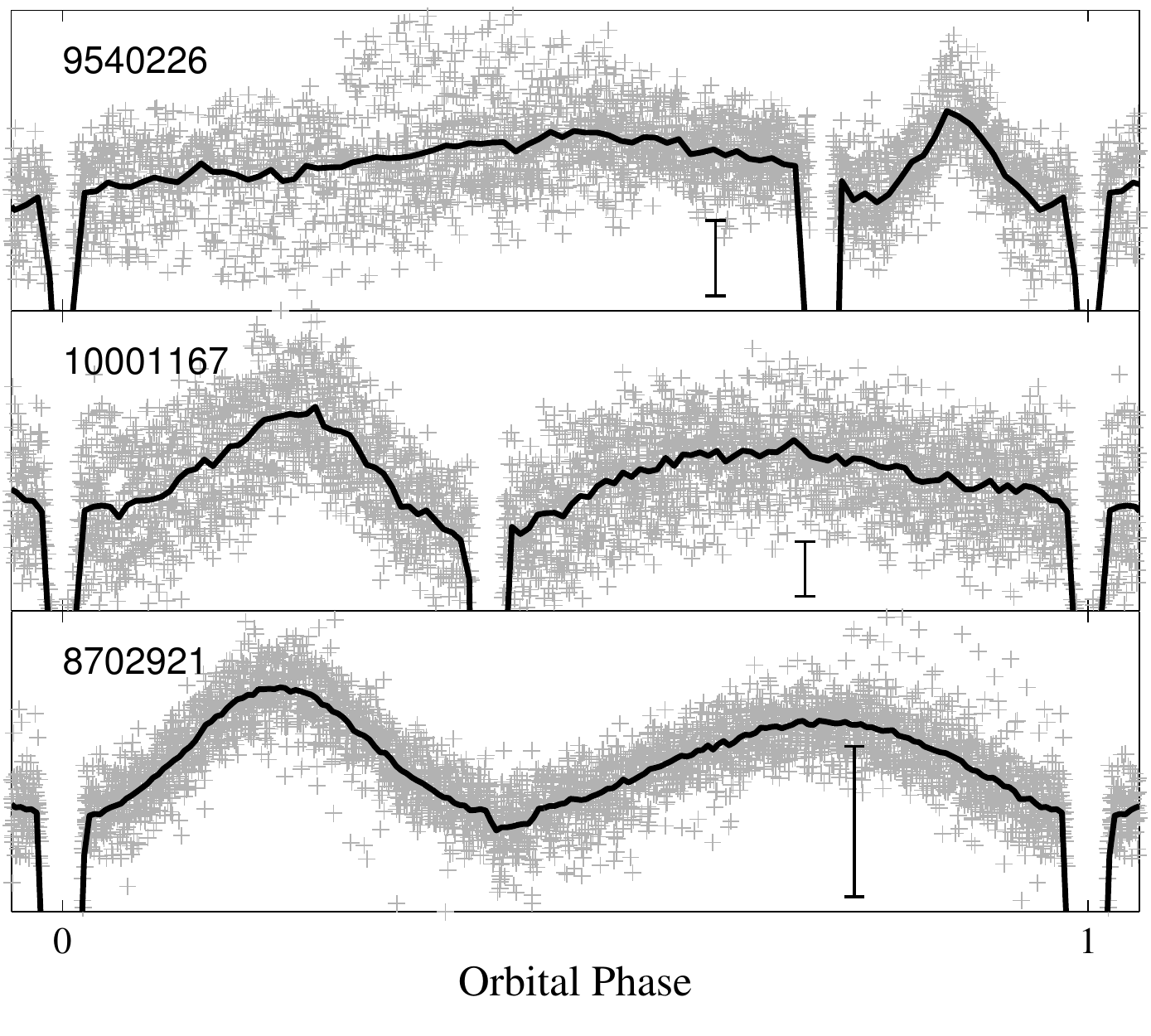}
\caption{Folded light curves of KICs 9540226, 10001167, and 8702921 as function of orbital phases from top to bottom. gray symbols represent the folded time series where 1 point out of 15 are displayed, and the thick black lines corresponds to the complete folded light curves rebinned in 2, 1, and 0.1-day steps respectively. The vertical bar corresponds to a relative flux variation of 0.2\,\%. Vertical axes are zoomed in to show evidence of the phase effects, while horizontal axes are set such as primary eclipses are in phases 0 and 1. \label{fig_phases} }
\end{center}
\end{figure}

In this paper, we utilized and extended the \citet{Gaulme_2013} light-curve processing, which preserves eclipse profiles, stellar variability, and oscillations. \citet{Gaulme_2013}  worked with ``PDCSAP\_FLUX'' which are corrected for  various instrumental trends, statistically determined by principal component analysis on a large sample of stars. However, some features such as deep eclipses may be erased or altered, as we observed for KIC 5640750, and long-term variabilities may be canceled too, as we observed for KICs 9246715, 9540750 and 10001167 (see Fig. \ref{fig_phases}). Here, we chose to directly work with the raw SAP\_FLUX, which are the fluxes integrated per mask aperture. This approach presents the advantage of controlling the data analysis and understanding its limits. 

The major challenge in concatenating light curves and studying stellar activity on periods longer than a quarter is to ensure a photometric continuity before and after each interruption. The first step is to start with an appropriate normalization of photometric fluxes per quarter. To base the normalization on stellar fluxes out of eclipses, we divide each quarter by the median of the light-curve in which all data corresponding to eclipses are removed. Then, as observed in most datasets, many jumps remain, often several times per quarter, and generally correspond to short interruptions of data acquisition. After carefully identifying all significant gaps, we process each jump in two possible ways. When a gap is short with respect to the photometric variability, if obvious, each side of the gap is fitted by a second order polynomial. Then, each polynomial fitting is extrapolated out to the middle of the gap and the difference between both extrapolated values is used to adjust the photometry. When a gap is longer than the variability period - this happened for KICs 4569590, 5308778, 3955867, 9291629, and 7943602 - it is not possible to use the previous method and we adjust the photometry with the difference of the means of each chunk surrounding the gap. Once the complete time series is leveraged and concatenated, a linear fitting is subtracted from it to compensate the decreasing instrumental sensitivity.

Since we work with raw data, instrumental trends are not removed. The only significant feature that we observe in the light curves with stable photometry (all systems with orbits longer than 110 days) is a periodic modulation corresponding to \textit{Kepler}'s orbit (372.5 days), with peak-to-peak amplitude ranging from 0.5\,\% to 3.8\,\%. This likely corresponds to the differential velocity aberration - the motion of the target across a fixed aperture smaller than the PSF - caused by the pixel scale breathing along the satellite's orbit. Because all time series have gaps, it is not possible to use Fourier filtering to remove this signal. We thus subtract from each light curve a 372.5-day period sine fitting and a first harmonic. This is enough to reduce the amplitude of these modulation to less than 0.5\,\%. Note that \citet{Beck_2013} reported to be unable to detect the 300-ppm Doppler beaming signal in the light curve of a heartbeat star, because the 95-day orbital period is too close to a quarter. Here, we do not consider such low signals and the variabilities we identify are all larger than 0.5\,\% (5000 ppm). 

The confidence in the quality of our data processing can be illustrated by the clear detection of orbital phase effects in three of our systems (Fig.~\ref{fig_phases}). KICs 9540226, 10001167, and 8702921 present evidence of ellipsoidal distortions of at least one of the stellar components. The peak-to-peak amplitude of their modulations are 0.25, 0.4, and 0.2\,\% and their orbital periods 175, 120, and 19 days respectively. Although we do not consider phase effects for scientific analysis in this paper, this demonstrates that we are able to monitor photometric variability down to 0.2\,\% over a broad range of periods. 

\section{Power Spectral Densities}
Fourier analysis is used to determine asteroseismic properties and variability periods. Eclipses must be accurately removed from each time series to avoid contamination of the PSD. We studied two alternative manners of removing eclipses from a light curve. A natural approach is to divide each light curve by a synthetic one obtained from eclipse modeling.  Two reasons make this approach inappropriate. Firstly, no fit being perfect, significant residuals remain during eclipse ingresses and egresses. Secondly, spot activity (Figs.~\ref{fig_RG_spots} \& \ref{fig_9246715}) is not canceled when dividing by a model, leaving therefore periodic features in the time series. As proposed by \citet{Gaulme_2013} for long-period systems, the most efficient way consists of bridging each eclipse with a second-order polynomial. The main drawback is a loss of signal, but it does not exceed 14.3\,\% (KIC 5308778) of the duty cycle. 

To minimize the effects of the incomplete duty cycle, we perform gap fillings and make use of the fast Fourier transform (FFT).  All short gaps (few missing points) are interpolated with a second order polynomial estimated from the nearby data points. Long gaps are filled with zeros. Besides, to reduce the impact of abrupt discontinuities around long gaps, the edges of each section in between gaps are apodized with a cosine function. This is particularly important when  significant activity is detected. 

The stellar variability for which fundamental periods are provided in the text are all well visible. First, proxies of their periods are obtained with autocorrelating the time series. Then, each measurement is refined by picking up the position of the highest peak of in the Fourier spectrum, at the corresponding frequency. In all our cases, the reported activity is so high that there are no doubts as to  which are the correct peaks in the PSD. The PSD is oversampled by a factor 10 to enable the best possible period estimate. 

As regards the search for oscillations, if no significant activity is present, no additional processing is done, and the FFT is computed. When activity is high, we filter it out with the help of a 15-hour long weighted moving average. Hence most of the signal with frequency lower than $20\ \mu$Hz is canceled. Note that the moving average is performed individually on each subsection in between large gaps. 
 
\acknowledgements


\bibliographystyle{apj}

\begin{thebibliography}{}


\bibitem[{{Auri{\`e}re} et~al.(2011){Auri{\`e}re}, {Konstantinova-Antova},
  {Petit} et~al.}]{Auriere_2011}
{Auri{\`e}re} M., {Konstantinova-Antova} R., {Petit} P., et~al., Oct. 2011,
  \aap, 534, A139
  
\bibitem[{{Baglin} {et~al.}(2009){Baglin}, {Auvergne}, {Barge}, {Deleuil},
  {Michel}, \& {The CoRoT Exoplanet Science Team}}]{Baglin_2009}
{Baglin}, A., {Auvergne}, M., {Barge}, P., {Deleuil}, M., {Michel}, E., \& {The
  CoRoT Exoplanet Science Team}. 2009, in IAU Symposium, Vol. 253, IAU
  Symposium, 71--81

\bibitem[{{Beck} et~al.(2013){Beck}, {Hambleton}, {Vos} et~al.}]{Beck_2013}
{Beck} P.G., {Hambleton} K., {Vos} J., et~al., Dec. 2013, ArXiv e-prints

\bibitem[{{Bonomo} \& {Lanza}(2012)}]{Bonomo_Lanza_2012}{Bonomo} A.S., {Lanza} A.F., Nov. 2012, \aap, 547, A37

\bibitem[{{Borucki} {et~al.}(2010){Borucki}, {Koch}, {Basri}, {Batalha},
  {Brown}, {Caldwell}, {Caldwell}, {Christensen-Dalsgaard}, {Cochran},
  {DeVore}, {Dunham}, {Dupree}, {Gautier}, {Geary}, {Gilliland}, {Gould},
  {Howell}, {Jenkins}, {Kondo}, {Latham}, {Marcy}, {Meibom}, {Kjeldsen},
  {Lissauer}, {Monet}, {Morrison}, {Sasselov}, {Tarter}, {Boss}, {Brownlee},
  {Owen}, {Buzasi}, {Charbonneau}, {Doyle}, {Fortney}, {Ford}, {Holman},
  {Seager}, {Steffen}, {Welsh}, {Rowe}, {Anderson}, {Buchhave}, {Ciardi},
  {Walkowicz}, {Sherry}, {Horch}, {Isaacson}, {Everett}, {Fischer}, {Torres},
  {Johnson}, {Endl}, {MacQueen}, {Bryson}, {Dotson}, {Haas}, {Kolodziejczak},
  {Van Cleve}, {Chandrasekaran}, {Twicken}, {Quintana}, {Clarke}, {Allen},
  {Li}, {Wu}, {Tenenbaum}, {Verner}, {Bruhweiler}, {Barnes}, \&
  {Prsa}}]{Borucki_2010}
{Borucki}, W.~J. {et~al.} 2010, Science, 327, 977


\bibitem[{{Chaplin} et~al.(2011){Chaplin}, {Bedding}, {Bonanno}
  et~al.}]{Chaplin_2011b}
{Chaplin} W.J., {Bedding} T.R., {Bonanno} A., et~al., May 2011, \apjl, 732, L5+

\bibitem[{{Derekas} {et~al.}(2011){Derekas}, {Kiss}, {Borkovits}, {Huber},
  {Lehmann}, {Southworth}, {Bedding}, {Balam}, {Hartmann}, {Hrudkova},
  {Ireland}, {Kov{\'a}cs}, {Mez{\H o}}, {Mo{\'o}r}, {Niemczura}, {Sarty},
  {Szab{\'o}}, {Szab{\'o}}, {Telting}, {Tkachenko}, {Uytterhoeven}, {Benk{\H
  o}}, {Bryson}, {Maestro}, {Simon}, {Stello}, {Schaefer}, {Aerts}, {ten
  Brummelaar}, {De Cat}, {McAlister}, {Maceroni}, {M{\'e}rand}, {Still},
  {Sturmann}, {Sturmann}, {Turner}, {Tuthill}, {Christensen-Dalsgaard},
  {Gilliland}, {Kjeldsen}, {Quintana}, {Tenenbaum}, \& {Twicken}}]{Derekas_2011}
{Derekas}, A. {et~al.} 2011, Science, 332, 216

\bibitem[{{Donati} \& {Brown}(1997)}]{Donati_Brown_1997}
{Donati} J., {Brown} S.F., Oct. 1997, \aap, 326, 1135

\bibitem[{{Dupree} \& {Smith}(1995)}]{Dupree_Smith_1995}
{Dupree} A.K., {Smith} G.H., Jul. 1995, \aj, 110, 405

\bibitem[{{Frandsen} et~al.(2013){Frandsen}, {Lehmann}, {Hekker}
  et~al.}]{Frandsen_2013}
{Frandsen} S., {Lehmann} H., {Hekker} S., et~al., Jul. 2013, ArXiv e-prints

\bibitem[{{Fuller} et~al.(2013){Fuller}, {Derekas}, {Borkovits}
  et~al.}]{Fuller_2013}
{Fuller} J., {Derekas} A., {Borkovits} T., et~al., Mar. 2013, \mnras, 429, 2425

\bibitem[{{Gaulme} et~al.(2013){Gaulme}, {McKeever}, {Rawls}
  et~al.}]{Gaulme_2013}
{Gaulme} P., {McKeever} J., {Rawls} M.L., et~al., Apr. 2013, \apj, 767, 82

\bibitem[{{Gaulme}(2013)}]{Gaulme_2013b}
{Gaulme} P., Dec. 2013, In: {Shibahashi} H., {Lynas-Gray} A.E. (eds.)
  Astronomical Society of the Pacific Conference Series, vol. 479 of
  Astronomical Society of the Pacific Conference Series, 185

\bibitem[{{Hekker} {et~al.}(2010){Hekker}, {Debosscher}, {Huber}, {Hidas}, {De
  Ridder}, {Aerts}, {Stello}, {Bedding}, {Gilliland}, {Christensen-Dalsgaard},
  {Brown}, {Kjeldsen}, {Borucki}, {Koch}, {Jenkins}, {Van Winckel}, {Beck},
  {Blomme}, {Southworth}, {Pigulski}, {Chaplin}, {Elsworth}, {Stevens},
  {Dreizler}, {Kurtz}, {Maceroni}, {Cardini}, {Derekas}, \&
  {Suran}}]{Hekker_2010}
{Hekker}, S. {et~al.} 2010, \apjl, 713, L187

\bibitem[{Huber {et~al.}(2011)Huber, Bedding, Stello, Hekker, Mathur, Mosser,
  Verner, Bonanno, Buzasi, Campante, Elsworth, Hale, Kallinger, Aguirre,
  Chaplin, Ridder, García, Appourchaux, Frandsen, Houdek, Molenda-Żakowicz,
  Monteiro, Christensen-Dalsgaard, Gilliland, Kawaler, Kjeldsen, Broomhall,
  Corsaro, Salabert, Sanderfer, Seader, \& Smith}]{Huber_2011}
Huber, D. {et~al.} 2011, The Astrophysical Journal, 743, 143

\bibitem[{{Huber} et~al.(2013){Huber}, {Silva Aguirre}, {Matthews}
  et~al.}]{Huber_2013}
{Huber} D., {Silva Aguirre} V., {Matthews} J.M., et~al., Dec. 2013, ArXiv
  e-prints
  
\bibitem[{{Konstantinova-Antova} et~al.(2012){Konstantinova-Antova},
  {Auri{\`e}re}, {Petit} et~al.}]{Konstantinova_2012}
{Konstantinova-Antova} R., {Auri{\`e}re} M., {Petit} P., et~al., May 2012,
  \aap, 541, A44

\bibitem[{{Konstantinova-Antova} et~al.(2013){Konstantinova-Antova},
  {Auri{\`e}re}, {Charbonnel} et~al.}]{Konstantinova_2013}
{Konstantinova-Antova} R., {Auri{\`e}re} M., {Charbonnel} C., et~al., Nov.
  2013, ArXiv e-prints

\bibitem[{{Mosser} et~al.(2009){Mosser}, {Baudin}, {Lanza}  et~al.}]{Mosser_Baudin_2009}{Mosser} B., {Baudin} F., {Lanza} A.F., et~al., Oct. 2009, \aap, 506, 245

\bibitem[{{Mosser} \& {Appourchaux}(2009)}]{Mosser_Appourchaux_2009}
{Mosser}, B., \& {Appourchaux}, T. 2009, \aap, 508, 877

\bibitem[{{Mosser} {et~al.}(2012){Mosser}, {Goupil}, {Belkacem},
  {Michel}, {Stello}, {Marques}, {Elsworth}, {Barban}, {Beck}, {Bedding}, {De
  Ridder}, {Garc{\'{\i}}a}, {Hekker}, {Kallinger}, {Samadi}, {Stumpe},
  {Barclay}, \& {Burke}}]{Mosser_2012}
---. 2012{\natexlab{}}, \aap, 540, A143

\bibitem[{{Mosser} et~al.(2013){Mosser}, {Michel}, {Belkacem}
  et~al.}]{Mosser_2013}
{Mosser} B., {Michel} E., {Belkacem} K., et~al., Feb. 2013, \aap, 550, A126

\bibitem[{{Ol{\'a}h} et~al.(2013){Ol{\'a}h}, {Mo{\'o}r}, {Strassmeier},
  {Borkovits}, \& {Granzer}}]{Olah_2013}
{Ol{\'a}h} K., {Mo{\'o}r} A., {Strassmeier} K.G., {Borkovits} T., {Granzer} T.,
  2013, Astronomische Nachrichten, 334, 625
  
  \bibitem[{{{\"O}zdarcan} et~al.(2010){{\"O}zdarcan}, {Evren}, {Strassmeier},
  {Granzer}, \& {Henry}}]{Ozdarcan_2010}
{{\"O}zdarcan} O., {Evren} S., {Strassmeier} K.G., {Granzer} T., {Henry} G.W.,
  Aug. 2010, Astronomische Nachrichten, 331, 794

\bibitem[{{Peirce}(1852)}]{Peirce_1852}
{Peirce} B., Jul. 1852, \aj, {\bf 2}, 161

\bibitem[{{Siess} \& {Livio}(1999)}]{Siess_Livio_1999}
{Siess} L., {Livio} M., Oct. 1999, \mnras, 308, 1133

\bibitem[{{Silva-Valio} et~al.(2010){Silva-Valio}, {Lanza}, {Alonso}, \&
  {Barge}}]{Silva-Valio_2010}
{Silva-Valio} A., {Lanza} A.F., {Alonso} R., {Barge} P., Feb. 2010, \aap, 510,
  A25

\bibitem[{{Simon} \& {Drake}(1989)}]{Simon_Drake_1989}
{Simon} T., {Drake} S.A., Nov. 1989, \apj, 346, 303
  
\bibitem[{{Slawson} {et~al.}(2011){Slawson}, {Pr{\v s}a}, {Welsh}, {Orosz},
  {Rucker}, {Batalha}, {Doyle}, {Engle}, {Conroy}, {Coughlin}, {Gregg},
  {Fetherolf}, {Short}, {Windmiller}, {Fabrycky}, {Howell}, {Jenkins}, {Uddin},
  {Mullally}, {Seader}, {Thompson}, {Sanderfer}, {Borucki}, \&
  {Koch}}]{Slawson_2011}
{Slawson}, R.~W. {et~al.} 2011, \aj, 142, 160

\bibitem[{{Southworth} {et~al.}(2009){Southworth}, {Hinse}, {Dominik},
  {Glitrup}, {J{\o}rgensen}, {Liebig}, {Mathiasen}, {Anderson}, {Bozza},
  {Browne}, {Burgdorf}, {Calchi Novati}, {Dreizler}, {Finet}, {Harps{\o}e},
  {Hessman}, {Hundertmark}, {Maier}, {Mancini}, {Maxted}, {Rahvar}, {Ricci},
  {Scarpetta}, {Skottfelt}, {Snodgrass}, {Surdej}, \&
  {Zimmer}}]{Southworth_2009}
{Southworth}, J. {et~al.} 2009, \apj, 707, 167

\bibitem[{{Stepien}(1993)}]{Stepien_1993}
{Stepien} K., Oct. 1993, \apj, 416, 368


\bibitem[{{Welsh} et~al.(2011){Welsh}, {Orosz}, {Aerts} et~al.}]{Welsh_2011}
{Welsh} W.F., {Orosz} J.A., {Aerts} C., et~al., Nov. 2011, \apjs, 197, 4

\end{thebibliography}

\end{document}